\documentclass[aps,prl,twocolumn,superscriptaddress,floatfix,nobalancelastpage]{revtex4-2}
\usepackage{graphicx}
\usepackage{amsmath,amssymb,amsfonts,bm}
\usepackage{hyperref}
\usepackage{xcolor}
\usepackage{changes}

\begin{document}

\title{Statistically Steady Holographic Quantum Turbulence:\\ Hyperuniform Vortex Matter and Crossover}

\author{Yu-Ping An}%
\email[ ]{yuping.an@campus.technion.ac.il}
\affiliation{Department of Physics, Technion, Haifa 32000, Israel}
\author{Peng-Bo Ding}
\email[ ]{dingpengbo@itp.ac.cn}
 \affiliation{Institute of Theoretical Physics,
Chinese Academy of Sciences, Beijing 100190, China}
\affiliation{School of Physical Sciences, University of Chinese Academy of Sciences, Beijing 100049, China}
\author{Zhen-Han Jin}
\email[ ]{jinzhenhan@itp.ac.cn}
 \affiliation{Institute of Theoretical Physics,
Chinese Academy of Sciences, Beijing 100190, China}
\affiliation{School of Physical Sciences, University of Chinese Academy of Sciences, Beijing 100049, China}
\author{Li Li}
\email[ ]{liliphy@itp.ac.cn}
 \affiliation{Institute of Theoretical Physics,
Chinese Academy of Sciences, Beijing 100190, China}
\affiliation{School of Physical Sciences, University of Chinese Academy of Sciences, Beijing 100049, China}
\affiliation{School of Fundamental Physics and Mathematical Sciences, Hangzhou Institute for Advanced Study, University of Chinese Academy of Sciences, Hangzhou 310024, China}


\begin{abstract}
A long-standing obstacle in quantum turbulence has been the difficulty of sustaining robust statistical steady states, preventing unambiguous identification of universal vortex organization and kinetic scaling. We construct such a steady state in two-dimensional holographic superfluid turbulence by continuous Landau-instability driving, sustaining ${\sim}2500$ vortices free from transient artifacts. The topological charge structure factor $S_c(k)$ reveals Class I disordered hyperuniformity with $S_c(k)\propto k^{\alpha>1}$ as $k\to0$, where $k$ is the wavenumber. This constitutes the strongest long-range order of its kind and its first observation in a strongly driven, far-from-equilibrium quantum fluid with topological defects as the organizing principle, establishing a novel non-equilibrium vortex phase. Exploiting this platform, we resolve the scaling controversy: the apparent $k^{-5/3}$ signature in the kinetic energy spectrum is a narrow crossover between the $k^{-1}$ single-vortex and $k^{-3}$ core regimes, not a genuine Kolmogorov inertial range. The real-space second-order structure function provides decisive evidence via $S_2(r)\propto \ln r$ where $r$ is the spatial separation, with no $r^{2/3}$ Kolmogorov scaling, ruling out a true inertial cascade. These findings reveal that strongly coupled quantum turbulence lacks an inverse cascade due to the absence of macroscopic Onsager clusters, demonstrating energy transport fundamentally distinct from weakly coupled superfluids.
\end{abstract}

\maketitle

\textit{Introduction.---}Defining emergent phases of matter and universal scaling laws in driven-dissipative quantum systems is a fundamental pursuit of non-equilibrium physics. Quantum turbulence—the chaotic, multi-scale flow of quantized vortices—provides an ideal testing ground for these principles~\cite{vinen2002quantum,barenghi2023quantum,boffetta2012two,barenghi2014introduction}. However, extracting intrinsic statistical properties requires a robust steady state; freely decaying setups inevitably entangle genuine scaling laws with transient decay dynamics and a continuously depleting vortex population, rendering conclusive identification of universal behavior problematic.

This fundamental ambiguity lies at the heart of a long-standing debate over the kinetic energy spectrum in two-dimensional (2D) quantum turbulence. In weakly coupled superfluids described by the Gross-Pitaevskii equation (GPE), a classical Kolmogorov \(k^{-5/3}\) inverse energy cascade emerges when same-sign vortices aggregate into macroscopic Onsager clusters~\cite{onsager1949statistical,bradley2012energy,reeves2013inverse,billam2015spectral,gauthier2019giant,johnstone2019evolution}. At strong coupling, perturbative methods fail, and holographic duality provides a non-perturbative, remarkably powerful alternative by recasting the strongly coupled field theory as classical gravity in one higher dimension~\cite{zaanen2015holographic,hartnoll2018holographic,liu2020quantum}. This framework has successfully addressed diverse non-equilibrium phenomena in superfluids, ranging from vortex dynamics~\cite{wittmer2021vortex,yang2023holographic,an2025stability,lan2023heating}, quantum turbulence~\cite{chesler2013holographic,lan2016towards,zeng2025dissipation}, Kibble-Zurek mechanism~\cite{xia2026kibble,chesler2015defect}, interface and bubble dynamics~\cite{an2024interface,an2024quantum,jin2026bubble}. Early holographic simulations of freely decaying 2D turbulence reported an apparent \(k^{-5/3}\) scaling~\cite{chesler2013holographic,lan2016towards}, yet its physical origin remains deeply controversial. Fundamentally, quantum energy spectra are constrained by vortex geometry: isolated vortices produce a \(k^{-1}\) velocity tail, while the vortex core structure dictates a \(k^{-3}\) ultraviolet cutoff. Interpreting an intermediate \(k^{-5/3}\) signature as either a genuine inertial cascade or merely a narrow crossover thus requires exceptional scrutiny. Crucially, without large-scale Onsager clustering and relying solely on decaying dynamics that are sensitive to initial conditions~\cite{chesler2013holographic,lan2016towards,ewerz2015non}, previous studies could not settle this fundamental question. 

In this Letter, we move clearly beyond transient regimes by constructing a robust, statistically steady state of 2D holographic superfluid turbulence. By imposing a background superflow above the Landau critical threshold, we continuously inject topological defects via the Landau instability. Operating on a spatial grid far exceeding prior simulations, our system sustains a dense, interacting ensemble of approximately 2500 quantized vortices over an extended temporal window, enabling unambiguous statistical analysis free from initial-condition contamination.

Strikingly, this steady ensemble reveals a hidden structural phase of driven vortex matter. By evaluating the topological charge structure factor, we discover that the vortex distribution exhibits {Class I disordered hyperuniformity~\cite{torquato2018hyperuniform}. This state describes an isotropic, liquid-like disordered system in which large-scale charge density fluctuations are anomalously suppressed. Specifically, the vortex charge structure factor scales as \(S_c(k) \propto k^{\alpha}\) with \(\alpha > 1\) as \(k \to 0\), demonstrating the strongest form of hyperuniformity. This finding establishes that strongly coupled quantum turbulence is not merely chaotic, but represents a novel non-equilibrium phase of vortex matter endowed with hidden long-range topological order.

\begin{figure*}[htbp]
\centering
\includegraphics[width=0.35\textwidth]{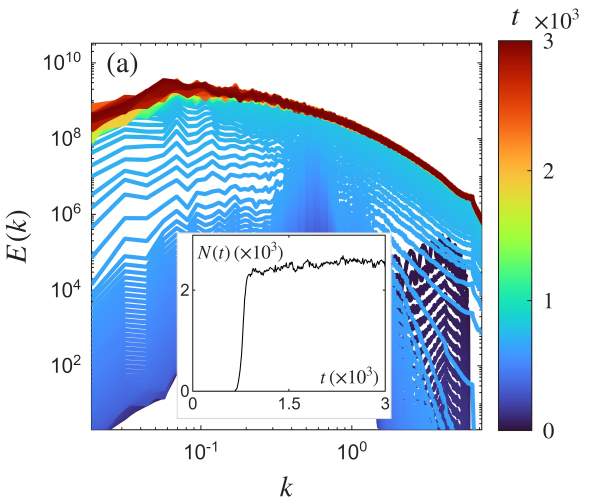}
\includegraphics[width=0.30\textwidth]{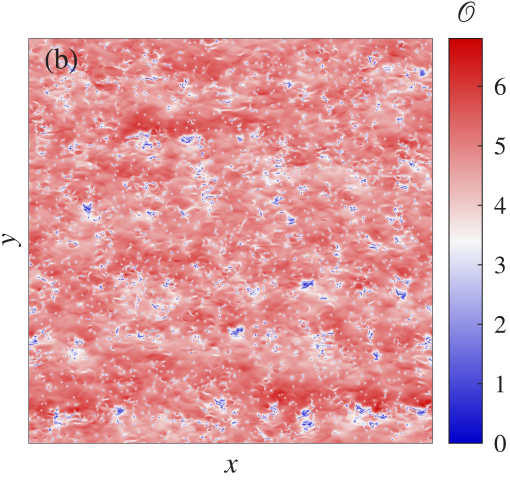}
\includegraphics[width=0.33\textwidth]{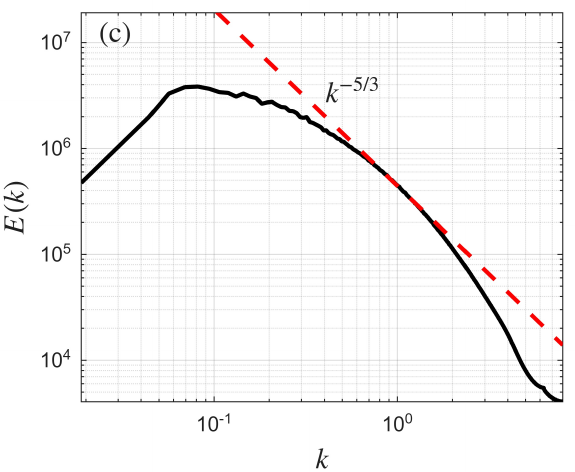}
\caption{Steady-state establishment and kinetic energy spectrum. (a) Time evolution of the kinetic energy spectrum and total vortex number $N(t)$ (inset), showing the system reaches a statistical steady state for $t\gtrsim 1000$ with $N\approx 2500$. (b) Snapshot of the order-parameter magnitude $\mathcal{O}$ at $t=3000$ over a $500\times500$ domain, revealing a dense, spatially homogeneous vortex ensemble with no macroscopic same-sign clusters. (c) Time-averaged total kinetic energy spectrum $E(k)$ computed from 200 snapshots in the steady-state window $t\in[1000,3000]$. An apparent narrow $k^{-5/3}$ region appears between the $k^{-1}$ single-vortex and $k^{-3}$ core regimes; its physical nature is scrutinized in Figs.~\ref{fig:structure_factor} and~\ref{fig:spectrum_average} below. Temperature  and superfluid velocity are fixed to $T/T_c=0.677$ and $v^s_y = 2.2$. All quantities are presented in units of $z_h$.
\label{fig:spectrum_tevol}}
\end{figure*}

Furthermore, this pristine steady state provides the definitive platform to resolve the kinetic scaling controversy. Through extensive time-averaging combined with dual-space (Fourier and real-space) statistics, we systematically isolate intrinsic collective dynamics from transient artifacts. Our energy spectra reveal that the apparent \(k^{-5/3}\) feature is not a broad inertial range but strictly a narrow crossover bridging the intermediate \(k^{-1}\) single-vortex regime and the short-distance \(k^{-3}\) core structure. To rigorously corroborate this finding, we compute the real-space second-order velocity structure function \(S_2(r)\). Our steady-state data exhibit robust \(S_2(r) \propto \ln r\) over an extended range of separations, which is the unique real-space signature of a \(k^{-1}\) spectrum. This logarithmic scaling constitutes decisive, parameter-free evidence that the observed \(k^{-5/3}\) behavior (corresponding to \(S_2(r) \propto r^{2/3}\)) is a crossover rather than a genuine inertial cascade.

Collectively, our results fundamentally redefine the asymptotic behaviors of strongly coupled 2D quantum turbulence. The absence of a true inverse cascade, rooted in the intrinsic dissipative vortex dynamics of holographic superfluids, reveals that energy transport in strongly coupled quantum fluids is governed by physics qualitatively distinct from the Onsager-clustering mechanism operative in weakly coupled systems. The discovery of disordered hyperuniformity in a driven-dissipative quantum fluid further opens a new avenue for classifying non-equilibrium phases of vortex matter. 

\textit{Setup and Steady-State Driving.---}We consider a strongly coupled $(2+1)$-dimensional superfluid described by the holographic Abelian-Higgs model in the probe limit~\cite{hartnoll2008building,herzog2009holographic}. The bulk action reads
\begin{equation}
S = \int d^4x \sqrt{-g} \left[ -\frac{1}{4}F_{\mu\nu}F^{\mu\nu} - |D \Psi|^2 - m^2|\Psi|^2 \right],
\end{equation}
where the complex scalar field $\Psi$ is minimally coupled to a $U(1)$ gauge field with $F_{\mu\nu}=\partial_\mu A_\nu-\partial_\nu A_\mu, D_\mu = \nabla_\mu - i A_\mu$. We work on a planar Schwarzschild–AdS$_4$ black hole background,
\begin{equation}
ds^2 = \frac{1}{z^2} \left( -f(z) dt^2 - 2 dt dz + dx^2 + dy^2 \right),
\end{equation}
where $f(z) = 1 - (z/z_h)^3$ with the temperature of the boundary fluid $T=3/(4\pi z_h)$. We set $m^2 = -2, z_h=1$ and choose the axial gauge $A_z=0$. Below the critical temperature $T_c$, the scalar $\Psi$ condenses, spontaneously breaking the $U(1)$ symmetry, and driving a second-order phase transition to a superfluid state with condensate $\langle \mathcal{O}\rangle=\mathcal{O} e^{i\theta}$. The superfluid velocity $v_i^s (i=x,y)$ can be obtained from the holographic dictionary. The details of the model are given in Supplemental Material (SM)~\cite{SM}.

To sustain a statistically steady turbulent state, we continuously inject energy and topological defects by imposing a constant background superflow. We set $T/T_c=0.677$ and fix the superfluid velocity $v^s_y = 2.2$, which exceeds the Landau critical threshold and triggers the Landau instability~\cite{amado2014holographic,lan2025landau}. This instability continuously nucleates vortex–antivortex pairs. We simulate on a large periodic domain of linear size $L_x=L_y=500$ (in units of $z_h$) with grid spacing well below the vortex core scale, ensuring faithful resolution of all relevant length scales (see Sec.\,S1 of SM for full numerical implementations). This setup vastly exceeds the spatial extent of previous holographic turbulence simulations. The dynamic balance between continuous vortex nucleation via the Landau instability and vortex-antivortex annihilation, together with dissipation into the black hole horizon, maintains a statistically steady vortex population without external tuning.

As shown in Fig.~\ref{fig:spectrum_tevol}(a) and its inset, after an initial transient the system settles into a non-equilibrium steady state sustaining $\sim $ 2500 quantized vortices. Crucially, we observe no formation of macroscopic same-sign vortex clusters (see Fig.~\ref{fig:spectrum_tevol}(b)), which is in sharp contrast to GPE-based quantum turbulence where sustained driving leads to Onsager clustering. This absence, rooted in the intrinsic dissipative vortex dynamics of holographic superfluids, already hints that energy transport in strongly coupled quantum fluids is qualitatively distinct from that in weakly coupled systems. Consequently, the time-averaged energy spectrum (Fig.~\ref{fig:spectrum_tevol}(c)) only shows a short $k^{-5/3}$ intermediate crossover. We have considered statistical homogeneity and isotropy~\cite{kolmogorov1991local,tsubota2017numerical}, which implies the energy spectra and structure factors below depend only on the amplitude $k$ of wave-vector $\mathbf{k}$~\footnote{The background superflow could introduce anisotropy at the driving scale. However, as demonstrated in SM Fig.\,S1, the directional anisotropy in both the spectral tensor and the structure function is negligible across the scales of interest, justifying the use of isotropic statistical measures. This is consistent with the general phenomenon of return to isotropy in turbulence in the literature.}.

\begin{figure}[t]
\centering
\includegraphics[width=0.46\textwidth]{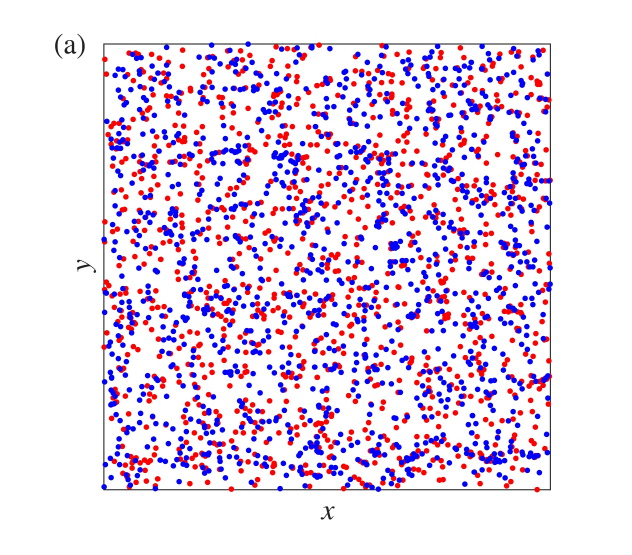}
\includegraphics[width=0.43\textwidth]{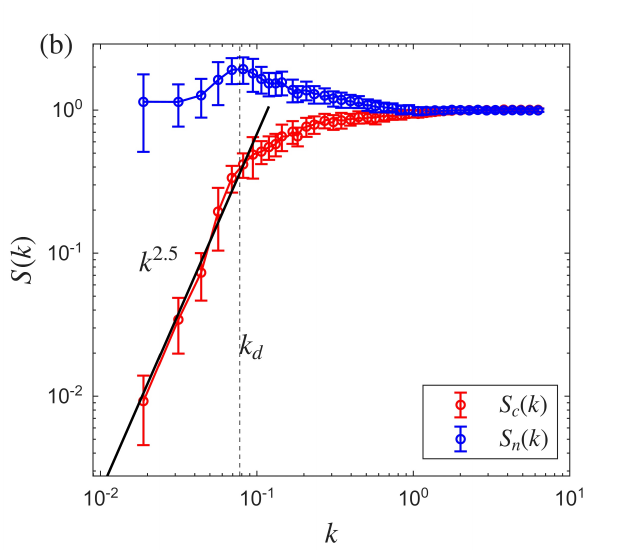}
\caption{Disordered hyperuniformity of the vortex charge distribution. (a) Snapshot of the vortex configuration at $t=3000$; red and blue markers denote positive and negative circulation, respectively. (b) Time-averaged structure factors. The number structure factor $S_n(k)$ (blue) does not vanish as $k\to0$, indicating no long-range number order. The charge structure factor $S_c(k)$ (red) exhibits a clear power-law suppression $S_c(k)\propto k^{\alpha}$ with $\alpha \approx 2.5$ as $k\to0$, demonstrating Class I disordered hyperuniformity. The peak in $S_n(k)$ at $k_d\simeq 0.08$ (vertical line) corresponds to a characteristic cluster spacing $d=2\pi/k_d\simeq 80$. Error bars denote standard deviations across 200 time slices. }
\label{fig:structure_factor}
\end{figure}

\textit{Vortex Structure Factor and Hyperuniformity.---}
We characterize the spatial organization of the steady-state vortex ensemble via the structure factor $S(k)$ for both vortex number and vortex charge~\cite{le2017enhanced,rumi2019hyperuniform,llorens2020disordered,sanchez2023disordered,milagros2024finite}. For a system of $N$ point vortices located at $\mathbf{r}_j$ with topological charges (i.e., winding numbers) $q_j = \pm 1$, we define the vortex number structure factor
$$S_n(\mathbf{k}) = \frac{1}{N} \left\langle \left| \sum_{j=1}^{N} e^{-i\mathbf{k} \cdot \mathbf{r}_j} \right|^2 \right\rangle$$
which probes the absolute spatial distribution of all vortices irrespective of circulation, and the vortex charge structure factor
$$S_c(\mathbf{k}) = \frac{1}{N} \left\langle \left| \sum_{j=1}^{N} q_j e^{-i\mathbf{k} \cdot \mathbf{r}_j} \right|^2 \right\rangle$$
which captures the spatial correlations of the net topological charge. These quantities are evaluated by time-averaging over 200 snapshots within the steady-state window $t \in [1000, 3000]$, ensuring statistical convergence.

Disordered hyperuniformity describes amorphous systems that simultaneously exhibit liquid-like isotropy and crystal-like suppression of long-wavelength density fluctuations~\cite{torquato2018hyperuniform}. In momentum space, this is signaled by an isotropic structure factor that vanishes as $k\rightarrow 0$, typically as a power law $S(k) \propto k^{\alpha}$. Hyperuniform systems are classified by the exponent $\alpha$: 
Class I $(\alpha > 1)$ exhibits the strongest suppression of large-scale fluctuations; 
Class II $(\alpha = 1)$ displays logarithmic corrections in real space; and Class III $(0 < \alpha < 1)$, represents the weakest form of hyperuniformity. Our results, shown in Fig.~\ref{fig:structure_factor}, reveal a striking dichotomy. The number structure factor $S_n(k)$ does not vanish as $k\rightarrow 0$,  indicating that the absolute vortex positions remain spatially disordered without long-range number correlations. Remarkably, however, the charge structure factor $S_c(k)$ exhibits a clear power-law suppression toward zero with $\alpha\approx 2.5$ in the low-wavenumber regime. This conclusively demonstrates Class I disordered hyperuniformity in the topological charge distribution—the strongest known form of hyperuniform order. Crucially, this hyperuniform scaling for $S_c(k)$ with $\alpha\approx 2.5$ is robust and independent of temperature variations (see Fig.\,S3 in SM).

\begin{figure*}[ht]
\centering
\includegraphics[width=1.05\textwidth]{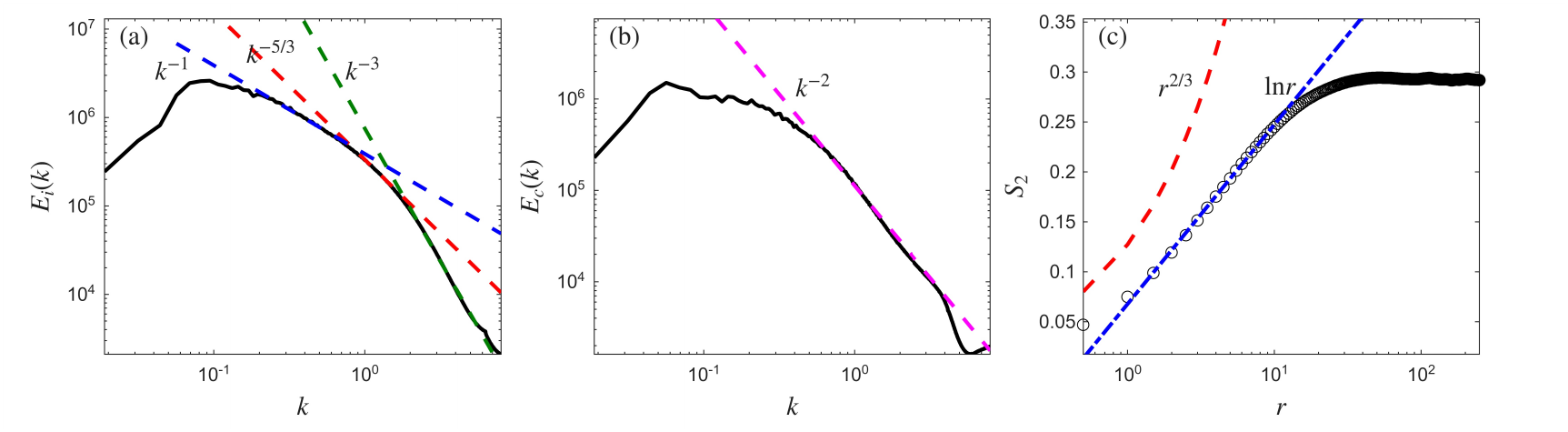}
\caption{Dual-space spectral analysis resolves the scaling controversy. (a) Time-averaged incompressible kinetic energy spectrum $E_i(k)$, showing the $k^{-1}$ isolated-vortex regime, the narrow $k^{-5/3}$ crossover, and the $k^{-3}$ core-scale regime. (b) Time-averaged compressible spectrum $E_c(k)$, which follows a $k^{-2}$ scaling due to shock-like acoustic waves from vortex--antivortex annihilations. (c) Second-order velocity structure function $S_2(r)$. This semi-log plot reveals a robust linear dependence, establishing $S_2(r)\propto \ln r$ (blue dashed)---the unique real-space signature of a $k^{-1}$ spectrum. This constitutes knockout, parameter-free evidence that the apparent $k^{-5/3}$ feature (corresponding to $S_2\propto r^{2/3}$) is a crossover.}
\label{fig:spectrum_average}
\end{figure*}

The peak in $S_n(k)$ at $k_d \simeq 0.08$ corresponds to a characteristic spacing $d = 2\pi/k_d \simeq 80$ between high-density vortex clusters identified via the Density-Peak Clustering algorithm~\cite{rodriguez2014clustering} (see SM Sec.~S3). Unlike Onsager clusters, these formations carry no net winding number and thus leave no signature in $S_c(k)$, consistent with the absence of macroscopic same-sign vortex aggregation. This structural distinction between number-density modulations and topological charge correlations is central to understanding why the inverse cascade prerequisite---large-scale same-sign vortex aggregation---is absent in the strongly coupled regime.

This finding carries profound physical implications. It reveals that despite the apparent chaotic motion of individual vortices, the interplay between positive and negative topological charges in this driven-dissipative system imposes a hidden long-range structural order that actively suppresses net charge fluctuations on arbitrarily large scales. Hyperuniform vortex configurations have been predicted in classical point-vortex systems~\cite{oppenheimer2022hyperuniformity} and observed in equilibrium type-II superconductor vortex lattices~\cite{le2017enhanced,rumi2019hyperuniform,llorens2020disordered,sanchez2023disordered,milagros2024finite}. Our work establishes, for the first time, hyperuniformity in a strongly driven, far-from-equilibrium quantum fluid with topological defects as the organizing principle. This regime is fundamentally distinct from both equilibrium vortex matter and classical point-vortex systems. 

\textit{Momentum and Real-Space Spectral Analysis.---} We now exploit the steady-state data to resolve the long-standing controversy over kinetic scaling laws. 
To separate the rotational vortex flow from the compressible acoustic density waves, we decompose the regularized velocity field as $\mathbf{v}^s = \mathbf{v}^s_i + \mathbf{v}^s_c$, where $\mathbf{v}^s_i$ and $\mathbf{v}^s_c$ are the incompressible and compressible parts, respectively. In Fourier space, the orthogonality of these transverse (incompressible) and longitudinal (compressible) modes gives the clean decomposition $E(k) = E_i(k) + E_c(k)$. 

The time-averaged energy spectra $E_i(k)$ and $E_c(k)$ in steady state are shown in Fig.~\ref{fig:spectrum_average}. The total spectrum $E(k)$ (Fig.~\ref{fig:spectrum_tevol}(c)) and its incompressible component $E_i(k)$ (Fig.~\ref{fig:spectrum_average}(a)) exhibit three distinct regimes. At intermediate wavenumbers, a robust $k^{-1}$ scaling emerges, reflecting the characteristic $v^s \sim 1/r$ velocity field of isolated quantized vortices. At large wavenumbers approaching the vortex core scale, a $k^{-3}$ scaling appears, capturing the intra-core profile of the superfluid order parameter. Between these two well-understood regimes lies a narrow transition region exhibiting an apparent $k^{-5/3}$ signature. Crucially, this feature spans barely half a decade in wavenumber space, which is strikingly narrower than the extended inertial range required for a genuine Kolmogorov cascade. The compressible spectrum $E_c(k)$ (Fig.~\ref{fig:spectrum_average}(b)) instead follows a $k^{-2}$ scaling, consistent with shock-like acoustic waves generated by frequent vortex--antivortex annihilations. This acoustic activity, intrinsic to the strongly coupled nature of the holographic superfluid, provides additional dissipation that actively suppresses large-scale vortex clustering.

The momentum-space observation of a narrow $k^{-5/3}$ feature, however, leaves open the possibility of finite-system or fitting ambiguities. We therefore turn to real-space statistics for an independent, parameter-free verification. The second-order velocity structure function $S_2(r)$ provides a direct probe of the velocity field's scale-dependent correlations without recourse to spectral fitting~\cite{davidson2015turbulence}. For a power-law energy spectrum $E(k) \propto k^{-n}$ with $1 < n < 3$, dimensional analysis dictates the real-space scaling $S_2(r) \propto r^{n-1}$. A genuine Kolmogorov inertial range with $n=5/3$ would therefore mandate $S_2(r) \propto r^{2/3}$. In stark contrast, an isolated-vortex-dominated $k^{-1}$ spectrum produces a logarithmic dependence $S_2(r) \propto \ln r$ (see Sec.\,S2 of SM).

Figure~\ref{fig:spectrum_average}(c) presents the steady-state time-averaged $S_2(r)$. The result completely fails to reveal any $r^{2/3}$ scaling region, conclusively ruling out the presence of a genuine inertial cascade. Conversely, the semi-log plot exhibits a robust linear dependence over more than a decade of spatial separations, establishing $S_2(r) \propto \ln r$ with high statistical confidence. This logarithmic scaling is not merely reminiscent of a \(k^{-1}\) spectrum; it behaves exactly as its unique real-space signature. The apparent $k^{-5/3}$ feature in Fourier space is thus unambiguously identified as a narrow crossover bridging the $k^{-1}$ single-vortex regime and the $k^{-3}$ core-scale regime. The structural origin of this crossover is addressed in the Discussion below. Our simulations of freely decaying setups (detailed in SM Sec.\,S5), by contrast, show that apparent $k^{-5/3}$ features are highly transient and strictly initial-condition dependent.

\textit{Discussion.---}
We have established a robust non-equilibrium steady state of 2D holographic superfluid turbulence, sustained by continuous Landau-instability driving. This platform eliminates the transient and initial-condition biases of all previous decaying studies, enabling unambiguous statistical analysis of a dense vortex ensemble (${\sim}2500$ vortices) over an extended temporal window.

Two interconnected results emerge. First, the vortex ensemble self-organizes into a state of Class I disordered hyperuniformity---the strongest form of hidden long-range order. The charge structure factor vanishes as $S_c(k) \propto k^\alpha$ with $\alpha \approx 2.5$, revealing that positive and negative topological charges cooperate to suppress net charge fluctuations on arbitrarily large scales. This constitutes the first observation of hyperuniformity in a driven-dissipative quantum fluid, identifying a novel phase of non-equilibrium vortex matter. Second, time-averaged spectral and real-space statistics conclusively resolve the long-standing scaling controversy: the apparent $k^{-5/3}$ signature is not a genuine Kolmogorov inertial range, but a narrow crossover bridging the $k^{-1}$ single-vortex and $k^{-3}$ core regimes. The real-space structure function provides decisive, parameter-free evidence: $S_2(r)\propto \ln r$, with no trace of the $r^{2/3}$ scaling mandated by a true cascade.

These results form a unified physical picture. The disordered hyperuniformity of the topological charge distribution (operating at \(k\lesssim 0.1\)) reveals a long-range hidden order that actively suppresses charge fluctuations on large scales. This structural constraint removes the physical prerequisite for an inverse cascade: in the absence of macroscopic same-sign clusters, the hierarchical vortex aggregation mechanism that drives the \(k^{-5/3}\) cascade in weakly coupled systems~\cite{bradley2012energy,reeves2013inverse,billam2015spectral,gauthier2019giant,johnstone2019evolution} cannot operate. Consequently, the dynamics at smaller scales (\(k\gtrsim 0.1\)) inevitably exhibit the observed narrow crossover between the \(k^{-1}\) single-vortex and \(k^{-3}\) core regimes, rather than a broad inertial range. The \(k^{-5/3}\) signature is therefore not an independent spectral feature but a direct consequence of the hyperuniform ordering principle across scales. 

From a broader perspective, our work opens several compelling frontiers. The robustness and controllability of our steady-state driving scheme provide an ideal platform for investigating transport properties of non-equilibrium strongly coupled superfluids, including vortex diffusion coefficients, effective turbulent viscosity, and the scale-by-scale energy flux budgets that remain challenging to compute due to compressible acoustic contributions. Extending this framework to fully back-reacted case and to 3D quantum turbulence, where vortex lines rather than point vortices carry topological charge, constitutes a natural and exciting direction for future research. Finally, the discovery of hyperuniformity in a driven quantum fluid suggests the possibility of a broader classification scheme for non-equilibrium phases of vortex matter, potentially bridging the gap between quantum turbulence, active matter, and the statistical mechanics of topological defects.

\begin{acknowledgments}
This work was supported by the National Natural Science Foundation of China Grants No.\,12525503, No.\,12588101 and No.\,12447101. We acknowledge the use of the High Performance Cluster at the Institute of Theoretical Physics, Chinese Academy of Sciences.
\end{acknowledgments}

\bibliography{biblio.bib}

@article{sanchez2023disordered,
  title={Disordered hyperuniform vortex matter with rhombic distortions in FeSe at low fields},
  author={S{\'a}nchez, Jazm{\'\i}n Arag{\'o}n and Maldonado, Ra{\'u}l Cort{\'e}s and Amig{\'o}, M Lourdes and Nieva, Gladys and Kolton, Alejandro and Fasano, Yanina},
  journal={Physical Review B},
  volume={107},
  number={9},
  pages={094508},
  year={2023},
  publisher={APS}
}

@article{milagros2024finite,
  title={Finite-size effects in hyperuniform vortex matter},
  author={Milagros Besana, Roc{\'\i}o and El{\'\i}as, Federico and Puig, Joaqu{\'\i}n and Arag{\'o}n S{\'a}nchez, Jazm{\'\i}n and Nieva, Gladys and Benedykt Kolton, Alejandro and Fasano, Yanina},
  journal={Journal of Physics: Condensed Matter},
  volume={36},
  number={28},
  pages={285102},
  year={2024},
  publisher={IOP Publishing}
}

@article{tsubota2017numerical,
  title={Numerical studies of quantum turbulence},
  author={Tsubota, Makoto and Fujimoto, Kazuya and Yui, Satoshi},
  journal={Journal of Low Temperature Physics},
  volume={188},
  number={5},
  pages={119--189},
  year={2017},
  publisher={Springer}
}

@article{liu2020quantum,
  title={Quantum many-body physics from a gravitational lens},
  author={Liu, Hong and Sonner, Julian},
  journal={Nature Reviews Physics},
  volume={2},
  number={11},
  pages={615--633},
  year={2020},
  publisher={Nature Publishing Group UK London}
}

@article{kolmogorov1991local,
  title={The local structure of turbulence in incompressible viscous fluid for very large Reynolds numbers},
  author={Kolmogorov, Andrei Nikolaevich},
  journal={Proceedings: Mathematical and Physical Sciences},
  volume={434},
  number={1890},
  pages={9--13},
  year={1991},
  publisher={JSTOR}
}

@article{rodriguez2014clustering,
  title={Clustering by fast search and find of density peaks},
  author={Rodriguez, Alex and Laio, Alessandro},
  journal={science},
  volume={344},
  number={6191},
  pages={1492--1496},
  year={2014},
  publisher={American Association for the Advancement of Science}
}

@article{oppenheimer2022hyperuniformity,
  title={Hyperuniformity and phase enrichment in vortex and rotor assemblies},
  author={Oppenheimer, Naomi and Stein, David B and Zion, Matan Yah Ben and Shelley, Michael J},
  journal={Nature communications},
  volume={13},
  number={1},
  pages={804},
  year={2022},
  publisher={Nature Publishing Group UK London}
}

@article{llorens2020disordered,
  title={Disordered hyperuniformity in superconducting vortex lattices},
  author={Llorens, Jos{\'e} Benito and Guillam{\'o}n, Isabel and Serrano, Ismael G and C{\'o}rdoba, Rosa and Ses{\'e}, Javier and De Teresa, Jos{\'e} Mar{\'\i}a and Ibarra, M Ricardo and Vieira, Sebasti{\'a}n and Ortu{\~n}o, Miguel and Suderow, Hermann},
  journal={Physical Review Research},
  volume={2},
  number={3},
  pages={033133},
  year={2020},
  publisher={APS}
}

@article{an2024quantum,
  title={Quantum analog to flapping of flags: interface instability for co-flow binary superfluids},
  author={An, Yu-Ping and Li, Li and Zeng, Hua-Bi},
  journal={Journal of High Energy Physics},
  volume={2024},
  number={10},
  pages={1--20},
  year={2024},
  publisher={Springer}
}

@article{zeng2025dissipation,
  title={Dissipation and decay of three-dimensional holographic quantum turbulence},
  author={Zeng, Hua-Bi and Xia, Chuan-Yin and Yang, Wei-Can and Tian, Yu and Tsubota, Makoto},
  journal={Physical Review Letters},
  volume={134},
  number={9},
  pages={091603},
  year={2025},
  publisher={APS}
}

@article{chesler2015defect,
  title={Defect formation beyond Kibble-Zurek mechanism and holography},
  author={Chesler, Paul M and Garcia-Garcia, Antonio M and Liu, Hong},
  journal={Physical Review X},
  volume={5},
  number={2},
  pages={021015},
  year={2015},
  publisher={APS}
}

@article{an2025stability,
  title={(In) stability of symbiotic vortex-bright soliton in holographic immiscible binary superfluids},
  author={An, Yu-Ping and Li, Li},
  journal={Journal of High Energy Physics},
  volume={2025},
  number={2},
  pages={42},
  year={2025},
  publisher={Springer}
}

@article{rumi2019hyperuniform,
  title={Hyperuniform vortex patterns at the surface of type-II superconductors},
  author={Rumi, Gonzalo and Arag{\'o}n S{\'a}nchez, Jazm{\'\i}n and El{\'\i}as, Federico and Cort{\'e}s Maldonado, Ra{\'u}l and Puig, Joaqu{\'\i}n and Cejas Bolecek, N{\'e}stor Ren{\'e} and Nieva, Gladys and Konczykowski, Marcin and Fasano, Yanina and Kolton, Alejandro B},
  journal={Physical Review Research},
  volume={1},
  number={3},
  pages={033057},
  year={2019},
  publisher={APS}
}

@article{le2017enhanced,
  title={Enhanced pinning for vortices in hyperuniform pinning arrays and emergent hyperuniform vortex configurations with quenched disorder},
  author={Le Thien, Quan and McDermott, D and Reichhardt, CJO and Reichhardt, C},
  journal={Physical Review B},
  volume={96},
  number={9},
  pages={094516},
  year={2017},
  publisher={APS}
}

@article{vinen2002quantum,
  title={Quantum turbulence},
  author={Vinen, William F and Niemela, Joseph J},
  journal={Journal of low temperature physics},
  volume={128},
  number={5},
  pages={167--231},
  year={2002},
  publisher={Springer}
}

@article{torquato2018hyperuniform,
  title={Hyperuniform states of matter},
  author={Torquato, Salvatore},
  journal={Physics Reports},
  volume={745},
  pages={1--95},
  year={2018},
  publisher={Elsevier}
}

@article{jin2026bubble,
  title={Bubble dynamics and vortex formation in holographic first-order superfluid phase transitions},
  author={Jin, Zhen-han and An, Yu-ping and Li, Li},
  journal={arXiv preprint arXiv:2604.17216},
  year={2026}
}

@book{hartnoll2018holographic,
  title={Holographic quantum matter},
  author={Hartnoll, Sean A and Lucas, Andrew and Sachdev, Subir},
  year={2018},
  publisher={MIT press}
}

@book{zaanen2015holographic,
  title={Holographic duality in condensed matter physics},
  author={Zaanen, Jan and Liu, Yan and Sun, Ya-Wen and Schalm, Koenraad},
  year={2015},
  publisher={Cambridge University Press}
}

@article{ewerz2015non,
  title={Non-thermal fixed point in a holographic superfluid},
  author={Ewerz, Carlo and Gasenzer, Thomas and Karl, Markus and Samberg, Andreas},
  journal={Journal of High Energy Physics},
  volume={2015},
  number={5},
  pages={1--36},
  year={2015},
  publisher={Springer}
}

@book{davidson2015turbulence,
  title={Turbulence: an introduction for scientists and engineers},
  author={Davidson, Peter},
  year={2015},
  publisher={Oxford university press}
}

@article{yang2024emergence,
  title={Emergence of large-scale structures in holographic superfluid turbulence},
  author={Yang, Wei-Can and Xia, Chuan-Yin and Tian, Yu and Tsubota, Makoto and Zeng, Hua-Bi},
  journal={arXiv preprint arXiv:2402.17980},
  year={2024}
}

@article{hartnoll2008building,
  title={Building a holographic superconductor},
  author={Hartnoll, Sean A and Herzog, Christopher P and Horowitz, Gary T},
  journal={Physical Review Letters},
  volume={101},
  number={3},
  pages={031601},
  year={2008},
  publisher={APS}
}

@article{chesler2013holographic,
  title={Holographic vortex liquids and superfluid turbulence},
  author={Chesler, Paul M and Liu, Hong and Adams, Allan},
  journal={Science},
  volume={341},
  number={6144},
  pages={368--372},
  year={2013},
  publisher={American Association for the Advancement of Science}
}

@article{lan2016towards,
  title={Towards quantum turbulence in finite temperature Bose-Einstein condensates},
  author={Lan, Shanquan and Tian, Yu and Zhang, Hongbao},
  journal={Journal of High Energy Physics},
  volume={2016},
  number={7},
  pages={1--16},
  year={2016},
  publisher={Springer}
}

@article{bradley2012energy,
  title={Energy spectra of vortex distributions in two-dimensional quantum turbulence},
  author={Bradley, Ashton S and Anderson, Brian P},
  journal={Physical Review X},
  volume={2},
  number={4},
  pages={041001},
  year={2012},
  publisher={APS}
}

@article{billam2015spectral,
  title={Spectral energy transport in two-dimensional quantum vortex dynamics},
  author={Billam, TP and Reeves, MT and Bradley, AS},
  journal={Physical Review A},
  volume={91},
  number={2},
  pages={023615},
  year={2015},
  publisher={APS}
}

@article{boffetta2012two,
  title={Two-dimensional turbulence},
  author={Boffetta, Guido and Ecke, Robert E},
  journal={Annual review of fluid mechanics},
  volume={44},
  number={1},
  pages={427--451},
  year={2012},
  publisher={Annual Reviews}
}

@article{onsager1949statistical,
  title={Statistical hydrodynamics},
  author={Onsager, Lars},
  journal={Il Nuovo Cimento (1943-1954)},
  volume={6},
  number={Suppl 2},
  pages={279--287},
  year={1949},
  publisher={Societ{\`a} Italiana di Fisica Bologna}
}

@article{reeves2013inverse,
  title={Inverse energy cascade in forced two-dimensional quantum turbulence},
  author={Reeves, Matthew T and Billam, Thomas P and Anderson, Brian P and Bradley, Ashton S},
  journal={Physical review letters},
  volume={110},
  number={10},
  pages={104501},
  year={2013},
  publisher={APS}
}

@article{barenghi2014introduction,
  title={Introduction to quantum turbulence},
  author={Barenghi, Carlo F and Skrbek, Ladislav and Sreenivasan, Katepalli R},
  journal={Proceedings of the National Academy of Sciences},
  volume={111},
  number={supplement\_1},
  pages={4647--4652},
  year={2014},
  publisher={National Academy of Sciences}
}

@article{gauthier2019giant,
  title={Giant vortex clusters in a two-dimensional quantum fluid},
  author={Gauthier, Guillaume and Reeves, Matthew T and Yu, Xiaoquan and Bradley, Ashton S and Baker, Mark A and Bell, Thomas A and Rubinsztein-Dunlop, Halina and Davis, Matthew J and Neely, Tyler W},
  journal={Science},
  volume={364},
  number={6447},
  pages={1264--1267},
  year={2019},
  publisher={American Association for the Advancement of Science}
}

@article{johnstone2019evolution,
  title={Evolution of large-scale flow from turbulence in a two-dimensional superfluid},
  author={Johnstone, Shaun P and Groszek, Andrew J and Starkey, Philip T and Billington, Christopher J and Simula, Tapio P and Helmerson, Kristian},
  journal={Science},
  volume={364},
  number={6447},
  pages={1267--1271},
  year={2019},
  publisher={American Association for the Advancement of Science}
}

@book{barenghi2023quantum,
  title={Quantum turbulence},
  author={Barenghi, Carlo F and Skrbek, Ladislav and Sreenivasan, Katepalli R},
  year={2023},
  publisher={Cambridge University Press}
}

@article{herzog2009holographic,
  title={Holographic model of superfluidity},
  author={Herzog, CP and Kovtun, PK and Son, DT},
  journal={Physical Review D—Particles, Fields, Gravitation, and Cosmology},
  volume={79},
  number={6},
  pages={066002},
  year={2009},
  publisher={APS}
}

@article{wittmer2021vortex,
  title={Vortex motion quantifies strong dissipation in a holographic superfluid},
  author={Wittmer, Paul and Schmied, Christian-Marcel and Gasenzer, Thomas and Ewerz, Carlo},
  journal={Physical Review Letters},
  volume={127},
  number={10},
  pages={101601},
  year={2021},
  publisher={APS}
}

@article{lan2023heating,
  title={Heating up quadruply quantized vortices: Splitting patterns and dynamical transitions},
  author={Lan, Shanquan and Li, Xin and Tian, Yu and Yang, Peng and Zhang, Hongbao},
  journal={arXiv preprint arXiv:2311.01316},
  year={2023}
}

@article{yang2023holographic,
  title={Holographic dissipative spacetime supersolids},
  author={Yang, Peng and Baggioli, Matteo and Cai, Zi and Tian, Yu and Zhang, Hongbao},
  journal={Physical Review Letters},
  volume={131},
  number={22},
  pages={221601},
  year={2023},
  publisher={APS}
}

@article{an2024interface,
  title={Interface dynamics of strongly interacting binary superfluids},
  author={An, Yu-Ping and Li, Li and Xia, Chuan-Yin and Zeng, Hua-Bi},
  journal={Physical Review D},
  volume={109},
  number={10},
  pages={106022},
  year={2024},
  publisher={APS}
}

@article{xia2026kibble,
  title={Kibble-Zurek mechanism and beyond in a holographic superfluid disk},
  author={Xia, Chuan-Yin and Zeng, Hua-Bi and Grabarits, Andr{\'a}s and Del Campo, Adolfo},
  journal={Nature Communications},
  year={2026},
  publisher={Nature Publishing Group UK London}
}

@article{amado2014holographic,
  title={Holographic superfluids and the Landau criterion},
  author={Amado, Irene and Are{\'a}n, Daniel and Jim{\'e}nez-Alba, Amadeo and Landsteiner, Karl and Melgar, Luis and Landea, Ignacio Salazar},
  journal={Journal of High Energy Physics},
  volume={2014},
  number={2},
  pages={1--24},
  year={2014},
  publisher={Springer}
}

@article{lan2025landau,
  title={Landau Instability and soliton formations},
  author={Lan, Shanquan and Liu, Hong and Tian, Yu and Zhang, Hongbao},
  journal={Physical Review D},
  volume={112},
  number={2},
  pages={L021901},
  year={2025},
  publisher={APS}
}

@misc{SM,
  note = {See Supplemental Material for detailed descriptions of the holographic setup, analytical derivation relating the kinetic energy spectrum in momentum space to the second-order structure function in real space, algorithm of identifying high-density vortex clusters, temperature robustness of hyperuniformity, and detailed analysis of freely decaying vortex turbulence demonstrating its dependence on initial conditions. It also includes~\cite{davidson2015turbulence,yang2024emergence}.}
}

\clearpage
\onecolumngrid
\begin{center}
  \textbf{\large Supplemental Material}\\[.2cm]
\end{center}

\vspace{0.3cm}
\setcounter{secnumdepth}{3}
\setcounter{section}{0}
\setcounter{equation}{0}
\setcounter{figure}{0}
\setcounter{table}{0}
\setcounter{page}{1}
\renewcommand{\theequation}{S\arabic{equation}}
\renewcommand{\thefigure}{S\arabic{figure}}
\renewcommand{\thetable}{S\arabic{table}}
\renewcommand{\thesection}{S\arabic{section}}

This Supplemental Material provides the full bulk equations of motion and numerical details (Section~\ref{app:eom}), the analytical derivation relating the kinetic energy spectrum in momentum space to the second-order structure function in real space (Section~\ref{app:relation}), the algorithm of identifying high-density vortex clusters (Section~\ref{app:dpc}), temperature dependence of hyperuniformity (Section~\ref{app:t_dependence}) and a detailed analysis of freely decaying vortex turbulence demonstrating its dependence on initial conditions (Section~\ref{app:decay}).

\section{Equations of motion and numerical details}\label{app:eom}

The equations of motion (EoMs) for $\Psi$ and $A_\mu$ are 
\begin{equation}
    D_\mu D^\mu \Psi - m^2 \Psi = 0\,,
\end{equation}
\begin{equation}
     \nabla_\mu F^{\mu\nu}=-2\mathrm{Im}(\Psi^*D^\nu\Psi)\,,
\end{equation}
where $\mathrm{Im}$ represents the imaginary part. Substituting into the Schwarzschild–AdS$_4$ black hole background,
\begin{equation}
ds^2 = \frac{1}{z^2} \left( -f(z) dt^2 - 2 dt dz + dx^2 + dy^2 \right),\quad f(z) = 1 - (z/z_h)^3\,,
\end{equation}
one has the explicit form of the EoMs:
    \begin{equation}
        \begin{aligned}
            \label{phi}
            2\partial_t\partial_z\Phi -[2i A_t\partial_z\Phi +i \partial_zA_t\Phi +\partial_z(f\partial_z\Phi )-z\Phi 
            +\partial_x^2\Phi +\partial_y^2\Phi 
            -i (\partial_xA_x+\partial_yA_y)\Phi &\\
            -(A_x^2+A_y^2)\Phi -2i (A_x\partial_x\Phi +A_y\partial_y\Phi )
            ]=0\,, &
        \end{aligned}
    \end{equation}
    \begin{equation}
        \label{At}
        \begin{aligned}
            \partial_t\partial_zA_t-[\partial_x^2A_t+\partial_y^2A_t+f\partial_z(\partial_xA_x+\partial_yA_y)-\partial_t(\partial_xA_x+\partial_yA_y)
            -2A_t  |\Phi |^2&\\
            -2f\mathrm{Im}(  \Phi ^*\partial_z\Phi )+2\mathrm{Im}(  \Phi ^*\partial_t\Phi )]=0\,,&
        \end{aligned}
    \end{equation}
    \begin{equation}
        \label{Ax}
        \begin{aligned}
            2\partial_t\partial_zA_x-[\partial_z(\partial_xA_t+f\partial_zA_x)+\partial_y(\partial_yA_x-\partial_xA_y)-2A_x  |\Phi |^2
            +2\mathrm{Im}(  \Phi ^*\partial_x\Phi )]=0\,,
        \end{aligned}
    \end{equation}
    \begin{equation}
        \label{Ay}
        \begin{aligned}           
        2\partial_t\partial_zA_y-[\partial_z(\partial_yA_t+f\partial_zA_y)+\partial_x(\partial_xA_y-\partial_yA_x)-2A_y  |\Phi |^2
            +2\mathrm{Im}(  \Phi ^*\partial_y\Phi )]=0\,,
        \end{aligned}
    \end{equation}
        \begin{equation}
        \label{constraint}
        \begin{aligned}     
        \partial_z(\partial_xA_x+\partial_yA_y-\partial_zA_t)-2\mathrm{Im}(  \Phi ^*\partial_z\Phi )=0\,,
        \end{aligned}
    \end{equation}
where $\Phi =\Psi /z$ together with the axial gauge $A_z=0$. We have highly nonlinear coupled PDEs for the fields $(\Phi, A_t, A_x, A_y)$ which are functions of $(t, z, x, y)$. Notice that the last equation is a constraint without time derivative. These equations are not independent. In particular,~\eqref{phi}, \eqref{Ax}, \eqref{Ay} and \eqref{constraint} guarantee that \eqref{At} is satisfied in the whole bulk as long as it is satisfied on any constant $z$ slice, such as on the AdS boundary $z=0$.

Near the AdS boundary $z \to 0$, the fields expand as
\begin{align}
\Psi &= z\Psi^{(1)} + z^2\Psi^{(2)} + O(z^3)\,, \\
A_\mu &= a_\mu + z b_\mu + O(z^2)\,,
\end{align}
where we set $m^2=-2$.
To drive spontaneous symmetry breaking, we set the vanishing source $\Psi^{(1)} = 0$. According to the holographic dictionary, the order parameter is $\langle \mathcal{O} \rangle = \mathcal{O} e^{i\theta}$, and $a_t = \mu$ is the chemical potential. The physical superfluid velocity is
\begin{equation}
v_s^i = \partial_i\theta - a_i (i=x,y)\,.
\end{equation}
To remove the unphysical $1/r$ divergence at the vortex cores when computing the momentum-space spectra, we use the density-weighted regularized velocity~\cite{chesler2013holographic,lan2016towards}
\begin{equation}
\mathbf{v}^s = \frac{(\nabla\theta - \mathbf{a})\mathcal{O}}{\max\{\mathcal{O}\}},
\end{equation}
with $\mathbf{a} = (a_x, a_y)$. 

Thanks to the scaling symmetry of EoMs, we set the horizon location $z_h=1$ in numerics. Then, by increasing $\mu$ we can effectively reduce the temperature to induce a superfluid phase transition. More precisely, the temperature on the boundary can be expressed as
\begin{equation}
T = T_c \frac{\mu_c}{\mu}\,,
\end{equation}
with the critical chemical potential $\mu_c=4.064$ in our current setup in the main text.

In this work, we drive the system by maintaining a global superfluid velocity $v^s_y=\partial_y\theta-a_y$. We set $T/T_c=0.677$ and fix the superfluid velocity $v^s_y = 2.2$, which exceeds the Landau critical threshold and triggers the Landau instability. In practice, to prevent an unphysical increase in global phase winding along $y$, we fix $a_y = -2.2$ (corresponding to $v^s_y = 2.2$) and dynamically monitor the spatial winding number,
\begin{equation}
W = \text{round} \left( \frac{1}{2\pi L_x} \int_0^{L_x} dx \int_0^{L_y} \partial_y \theta(x,y) dy \right)\,,
\end{equation}
where $L_x$ and $L_y$ are the box size of simulation region. Whenever $|W| \ge 1$, we apply a discrete phase shift $\Psi \to \Psi e^{-i 2\pi W y / L_y}$, which subtracts a uniform gradient and leaves velocity fluctuations unpolluted.

For time evolution, we use fourth order Runge-Kutta method with $dt=0.02$ and the following scheme: First, we use \eqref{phi}, \eqref{Ax} and \eqref{Ay} to evolve $\Phi$, $A_x$ and $A_y$ with boundary conditions $\Phi(z=0)=A_x(z=0)=A_y(z=0)=0$. Then we use \eqref{At} to evolve $\partial_zA_t$ on the boundary. Note that $-\partial_zA_t(z=0)$ is just the charge density or number density $\rho$ of the dual field theory. Finally, we use~\eqref{constraint} to solve $A_t$ by evolved $\Phi$, $A_x$, $A_y$ and boundary conditions $\partial_zA_t(z=0)=-\rho$ and $A_t(z=0)=\mu_{init}$. In the $z$ direction, we use the Chebyshev pseudo-spectral method with 20 grid points. In $x$ and $y$ directions we use the Fourier pseudo-spectral method with 1001 grid points and periodic boundary condition. The 3D box size of simulation region is $500\times500\times1$ in units of $z_h$. Due to periodic boundary conditions in spatial directions, total charge $Q=\int\rho\mathrm{d}x\mathrm{d}y$ is unchanged, so we are in fact working in the canonical ensemble.

\section{Relation between energy spectrum and second order structure function}\label{app:relation}

To rigorously analyze the energy distribution and distinguish the underlying physical processes, we perform a Helmholtz decomposition of the superfluid velocity field $\mathbf{v}^s$. Mathematically, any sufficiently smooth vector field can be decomposed into a divergence-free (solenoidal) component $\mathbf{v}^s_i$ and a curl-free (irrotational) component $\mathbf{v}^s_c$, such that $\mathbf{v}^s = \mathbf{v}^s_i + \mathbf{v}^s_c$, with $\nabla \cdot \mathbf{v}^s_i = 0$ and $\nabla \times \mathbf{v}^s_c = 0$. Physically, this separation is highly instructive for quantum turbulence: the incompressible component $\mathbf{v}^s_i$ predominantly captures the rotational flow induced by the quantized vortices, whereas the compressible component $\mathbf{v}^s_c$ isolates density perturbations, including the intense acoustic density waves and shock-like structures emitted during vortex-antivortex annihilations.

Crucially, this spatial decomposition naturally extends to an exact partitioning of the kinetic energy spectrum. The total kinetic energy spectrum $E(k)$ is obtained by integrating the velocity spectral density over the angular coordinate $\theta_k$ in momentum space:
\begin{equation}
E(k) = \frac{1}{2} \int |\tilde{\mathbf{v}}^s(\mathbf{k})|^2 k d\theta_k\,,
\end{equation}
where $\tilde{\mathbf{v}}^s(\mathbf{k})$ is the Fourier transform of the total velocity field. Similarly, one can define the kinetic energy spectrums $E_i(k)$ and $E_c(k)$ for incompressible and compressible parts, respectively. Substituting the Helmholtz decomposition $\tilde{\mathbf{v}}^s(\mathbf{k}) = \tilde{\mathbf{v}}^s_i(\mathbf{k}) + \tilde{\mathbf{v}}^s_c(\mathbf{k})$ into the integrand yields:
\begin{equation}
|\tilde{\mathbf{v}}^s(\mathbf{k})|^2 = |\tilde{\mathbf{v}}^s_i(\mathbf{k})|^2 + |\tilde{\mathbf{v}}^s_c(\mathbf{k})|^2 + \tilde{\mathbf{v}}^s_i(\mathbf{k}) \cdot [\tilde{\mathbf{v}}^s_c(\mathbf{k})]^* + [\tilde{\mathbf{v}}^s_i(\mathbf{k})]^* \cdot \tilde{\mathbf{v}}^s_c(\mathbf{k})\,.
\end{equation}
Because the incompressible component is divergence-free and the compressible component is curl-free in real space, their Fourier counterparts satisfy:
\begin{equation}
\mathbf{k} \cdot \tilde{\mathbf{v}}^s_i(\mathbf{k}) = 0, \quad \mathbf{k} \times \tilde{\mathbf{v}}^s_c(\mathbf{k}) = 0\,.
\end{equation}
This implies that $\tilde{\mathbf{v}}^s_i(\mathbf{k})$ is strictly transverse to the wave vector $\mathbf{k}$, while $\tilde{\mathbf{v}}^s_c(\mathbf{k})$ is strictly longitudinal (parallel to $\mathbf{k}$). Consequently, the two vector fields are perfectly orthogonal in momentum space, causing the cross-terms to vanish identically:
\begin{equation}
\tilde{\mathbf{v}}^s_i(\mathbf{k}) \cdot [\tilde{\mathbf{v}}^s_c(\mathbf{k})]^* = 0\,.
\end{equation}
As a result, the total kinetic energy spectrum is cleanly and exactly partitioned into an incompressible and a compressible part, i.e., $E(k) = E_i(k) + E_c(k)$. 

The second order structure function is defined as the statistical average of the squared velocity difference between two points separated by a spatial vector $\mathbf{r}$:
\begin{equation}\label{eqS2}
    S_2(r) = \langle |\mathbf{v}(\mathbf{x}+\mathbf{r}) - \mathbf{v}(\mathbf{x})|^2 \rangle\,.
\end{equation} 
It provides a direct probe
of the velocity field’s scale-dependent correlations without recourse to spectral fitting. 

In two-dimensional isotropic turbulence, the derivation of the relationship between the second order structure function $S_2(r)$ and the kinetic energy spectrum $E(k)$ proceeds as follows~\cite{davidson2015turbulence}.

First, expanding the square of~\eqref{eqS2} gives
\begin{equation}
    S_2(r) = \langle |\mathbf{v}(\mathbf{x}+\mathbf{r})|^2 \rangle + \langle |\mathbf{v}(\mathbf{x})|^2 \rangle - 2 \langle \mathbf{v}(\mathbf{x}+\mathbf{r}) \cdot \mathbf{v}(\mathbf{x}) \rangle\,.
\end{equation}
Assuming homogeneous and isotropic turbulence, the mean square velocity is uniform across the space, i.e., $\langle |\mathbf{v}(\mathbf{x}+\mathbf{r})|^2 \rangle = \langle |\mathbf{v}(\mathbf{x})|^2 \rangle = R(0)$, and the two-point velocity correlation function $R(r) \equiv \langle \mathbf{v}(\mathbf{x}+\mathbf{r}) \cdot \mathbf{v}(\mathbf{x}) \rangle$ depends solely on the scalar distance $r = |\mathbf{r}|$. This assumption is justified in Fig.~\ref{fig:isotropy} below.  Thus, the structure function simplifies to:
\begin{equation}\label{eq:S2_R}
    S_2(r) = 2 [R(0) - R(r)]\,.
\end{equation}

Second, in the Fourier space, the velocity correlation function $R(r)$ is closely related to the velocity spectral tensor $\Phi_{ij}(\mathbf{k})$:
\begin{equation}
    R(r) = \int \Phi_{ii}(\mathbf{k}) e^{i \mathbf{k} \cdot \mathbf{r}} \frac{d^2\mathbf{k}}{(2\pi)^2}\,,
\end{equation}
where $\Phi_{ii}(\mathbf{k})$ is the trace of $\Phi_{ij}(\mathbf{k})$, defined as the Fourier transform of the two-point velocity correlation tensor, i.e., $\Phi_{ij}(\mathbf{k}) = \int \langle v_i(\mathbf{x}+\mathbf{r})v_j(\mathbf{x}) \rangle e^{-i \mathbf{k} \cdot \mathbf{r}} \, d^2\mathbf{r}$.
In 2D polar coordinates, letting $\mathbf{k} = (k \cos\theta_k, k \sin\theta_k)$ and aligning $\mathbf{r}$ with the $x$-axis so that $\mathbf{k} \cdot \mathbf{r} = k r \cos\theta_k$, the momentum space area element becomes $d^2\mathbf{k} = k \, dk \, d\theta_k$. For isotropic turbulence, $\Phi_{ii}(\mathbf{k}) = \Phi_{ii}(k)$, yielding:
\begin{equation}
    R(r) = \int_0^\infty \frac{k \, dk}{(2\pi)^2} \Phi_{ii}(k) \int_0^{2\pi} e^{i k r \cos\theta_k} d\theta_k\,.
\end{equation}
Using the integral definition of the zeroth-order Bessel function of the first kind, $J_0(kr) = \frac{1}{2\pi} \int_0^{2\pi} e^{i k r \cos\theta} d\theta$, we obtain:
\begin{equation}\label{eq:R_Phi}
    R(r) = \int_0^\infty \frac{\Phi_{ii}(k)}{2\pi} J_0(kr) k \, dk\,.
\end{equation}

Third, the total turbulent kinetic energy per unit area is given by $\mathcal{K} = \frac{1}{2} \langle |\mathbf{v}|^2 \rangle = \frac{1}{2} R(0)$. The kinetic energy spectrum $E(k)$ is defined such that its integral over all wavenumbers gives the total kinetic energy:
\begin{equation}
    \mathcal{K} = \int_0^\infty E(k) \, dk\,.
\end{equation}
Evaluating $R(0)$ using~\eqref{eq:R_Phi} and noting that $J_0(0) = 1$, we have:
\begin{equation}
    \frac{1}{2} R(0) = \frac{1}{2} \int_0^\infty \frac{\Phi_{ii}(k)}{2\pi} k \, dk = \int_0^\infty \left( \frac{k \Phi_{ii}(k)}{4\pi} \right) dk\,.
\end{equation}
Comparing the two expressions for $\mathcal{K}$, we deduce the relation between the 2D energy spectrum and the spectral tensor:
\begin{equation}
    E(k) = \frac{k \Phi_{ii}(k)}{4\pi} \implies \frac{k \Phi_{ii}(k)}{2\pi} = 2 E(k)\,.
\end{equation}

Finally, substituting this relation back into the expression for $R(r)$ yields:
\begin{equation}
    R(r) = 2 \int_0^\infty E(k) J_0(kr) \, dk\,.
\end{equation}
Evaluating this correlation at $r=0$ gives $R(0) = 2 \int_0^\infty E(k) \, dk$. By substituting $R(0)$ and $R(r)$ into the structure function relation~\eqref{eq:S2_R}, we extract the common factor and arrive at the final result:
\begin{equation}
    S_2(r) = 4 \int_0^\infty E(k) [1 - J_0(kr)] \, dk\,.
\end{equation}

For a generic power-law energy spectrum $E(k) \propto k^{-n}$ within the range $1 < n < 3$, introducing the dimensionless integration variable $y = kr$ yields:
$$S_2(r) \propto \int_0^\infty \left(\frac{y}{r}\right)^{-n} [1 - J_0(y)] \frac{dy}{r} = r^{n-1} \int_0^\infty y^{-n} [1 - J_0(y)] \, dy\,.$$
Because the dimensionless integral over $y$ converges to a constant for $1 < n < 3$, the structure function scales algebraically as $S_2(r) \propto r^{n-1}$. Based on this derivation, a genuine Kolmogorov-like inertial range with $n=5/3$ dictates a real-space scaling of $S_2(r) \propto r^{2/3}$.

Conversely, for the marginal case of $n=1$ corresponding to the $E(k) \propto k^{-1}$ spectrum of isolated vortices, the aforementioned integral diverges. To evaluate it, one must introduce physical cutoffs: an infrared scale defined by the system size $L$, and an ultraviolet scale determined by the vortex core size $\xi$. Within the inertial range $\xi \ll r \ll L$, the term $[1 - J_0(kr)]$ acts effectively as a high-pass filter, vanishing for $k \ll 1/r$ and averaging to unity for $k \gg 1/r$. The dominant contribution to the integral thus comes from the interval $[1/r, 1/\xi]$, leading to:
$$S_2(r) \propto \int_{1/r}^{1/\xi} k^{-1} \, dk = \ln\left(\frac{r}{\xi}\right) \propto \ln r\,.$$
Therefore, the $k^{-1}$ energy spectrum manifests uniquely as a logarithmic dependence in real space. 

\begin{figure}[htbp]
\centering
\includegraphics[width=0.45\textwidth]{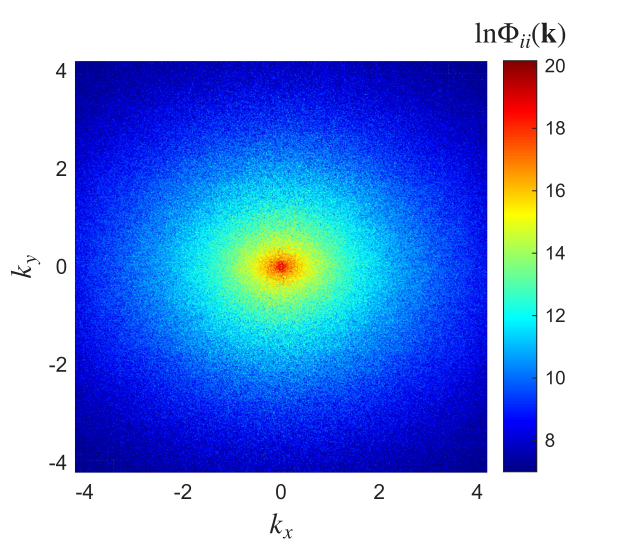}
\includegraphics[width=0.47\textwidth]{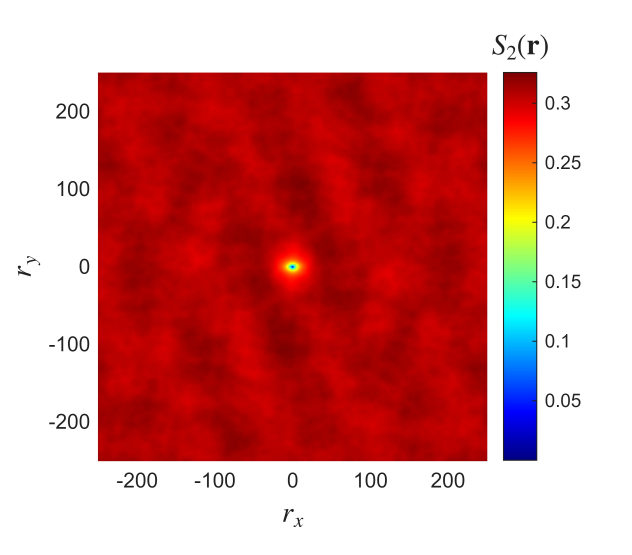}
\caption{\textbf{Left:} Trace of spectral tensor $\Phi_{ii}(\mathbf{k})$ at $t=3000$. \textbf{Right:} 2D structure function $S_2(\mathbf{r})$ at $t=3000$. These results directly demonstrate that anisotropy is negligible in $\Phi_{ii}(\mathbf{k})$ and $S_2(\mathbf{r})$, despite the background superflow.
\label{fig:isotropy}}
\end{figure}
In principle, the background superflow introduces anisotropy, requiring the spectral density and structure functions to depend explicitly on direction—i.e., $\Phi_{ii}(\mathbf{k})$ and $S_2(\mathbf{r})$ rather than their isotropic counterparts $\Phi_{ii}(k)$ and $S_2(r)$. However, while this large-scale driving breaks global rotational symmetry, directional memory fades at smaller spatial scales (higher wavenumbers). In line with Kolmogorov's local isotropy hypothesis~\cite{kolmogorov1991local}, strong nonlinear interactions and chaotic vortex dynamics progressively randomize the velocity fluctuations. This process dynamically decouples the short-range kinematics from the macroscopic flow, restoring rotational symmetry~\cite{tsubota2017numerical}. As directly demonstrated in Fig.~\ref{fig:isotropy}, the anisotropy in both $\Phi_{ii}(\mathbf{k})$ and $S_2(\mathbf{r})$ remains negligible across our scales of interest, thereby justifying the isotropic approximations $\Phi_{ii}(k)$ and $S_2(r)$.

\section{Identification of High-Density Vortex Clusters}
\label{app:dpc}

As shown in Fig.\,2(b) of the main text, the number structure factor $S_n(k)$ exhibits a prominent low-wavenumber peak corresponding to a characteristic spatial scale of approximately $80$, despite the system's microscopic inter-vortex spacing being much smaller ($\sim 10$). To establish the physical origin of this feature and verify whether it reflects macroscopic spatial modulations in the vortex number density, we employ a parameter-free spatial point-pattern analysis based on the Density-Peak Clustering (DPC) algorithm~\cite{rodriguez2014clustering}. 

Consider an ensemble of $N$ point vortices located at positions $\mathbf{r}_i = (x_i, y_i)$ within a simulation domain of dimensions $L_x \times L_y$. Denote the distance between any two vortices $i$ and $j$ as $d_{ij}$.
To eliminate discrete shot noise while preserving macroscopic density variations, the local vortex density $\rho_i$ at position $\mathbf{r}_i$ is evaluated using a continuous Gaussian smoothing kernel:
\begin{equation}
    \rho_i = \sum_{j \neq i}^{N} \exp\left( -\frac{d_{ij}^2}{d_c^2} \right),
\end{equation}
where the cutoff radius $d_c$ is chosen to match the microscopic inter-vortex scale to effectively capture local density modulations. Our system contains $\sim 2500$ vortices in a $L_x\times L_y=500\times 500$ domain, yielding a mean inter-vortex spacing $d_c \approx 10$, well below the characteristic cluster scale identified below.

For each vortex $i$, we subsequently compute $\delta_i$, defined as the minimum distance to any other vortex possessing a strictly higher local density:
\begin{equation}
    \delta_i = \begin{cases}
        \min_{j: \rho_j > \rho_i} d_{ij}, & \text{if } \exists \, j \text{ such that } \rho_j > \rho_i, \\
        \max_j d_{ij}, & \text{otherwise (global maximum density)}.
    \end{cases}
\end{equation}
Physically, vortices situated near local density peaks exhibit exceptionally large values of $\delta_i$, as crossing a lower-density valley is required to reach another vortex with a higher density.

High-density cluster centroids uniquely correspond to points possessing both a high local density $\rho_i$ and an anomalously large higher-density distance $\delta_i$. We construct the combined decision metric $\gamma_i = \rho_i \delta_i$. In the decision space spanned by $\rho$ and $\delta$, cluster centroids stand out as prominent statistical outliers. These centroids can be automatically extracted by enforcing a robust statistical threshold, such as $\gamma_{\text{th}} = \langle \gamma \rangle + 3 \sigma_\gamma$, where $\langle \gamma \rangle$ and $\sigma_\gamma$ denote the mean and standard deviation of $\gamma$ across the ensemble.

\begin{figure}[htbp]
\centering
\includegraphics[width=1\textwidth]{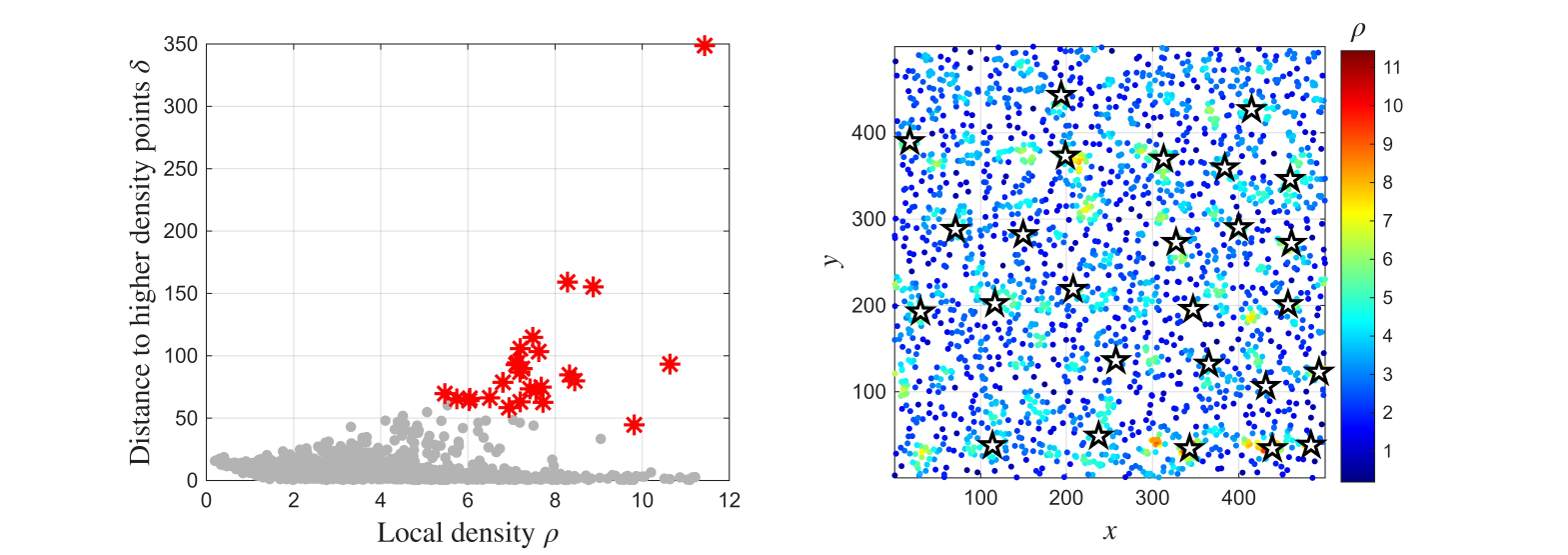}
\caption{Demonstration of the DPC algorithm at $t=3000$. \textbf{Left:} Decision Graph. Red stars at upper right denote centroids. \textbf{Right:} High-Density Cluster Centroids. Stars denote identified vortex clusters. 26 clusters are identified in this snapshot. Mean nearest-neighbor distance between these clusters is 74.26.
\label{fig:cluster}}
\end{figure}

Once the $N_c$ cluster centroids $\{\mathbf{r}_c^{(k)}\}_{k=1}^{N_c}$ are identified, the spatial distribution of these macroscopic density peaks is characterized by evaluating their mean nearest-neighbor distance:
\begin{equation}
    d = \frac{1}{N_c} \sum_{k=1}^{N_c} \min_{m \neq k} d_{m, k},
\end{equation}
where $d_{m, k}$ represents the distance between centroid $k$ and centroid $m$. In Fig.~\ref{fig:cluster}, we demonstrate this algorithm for a single time slice $t=3000$. $k_d$ in the main text is extracted by  time average. The direct measurement of $d \approx 80$ (after time average) independently confirms that the low-wavenumber peak observed in $S_n(k)$ originates from the characteristic real-space separation between macroscopic high-density vortex clusters. However, because these clusters consist primarily of vortex pairs and thus carry no net winding number, they are fundamentally distinct from Onsager clusters. Consequently, they leave no signature in the topological charge structure factor $S_c(k)$.

\begin{figure}[htbp]
\centering
\includegraphics[width=0.8\textwidth]{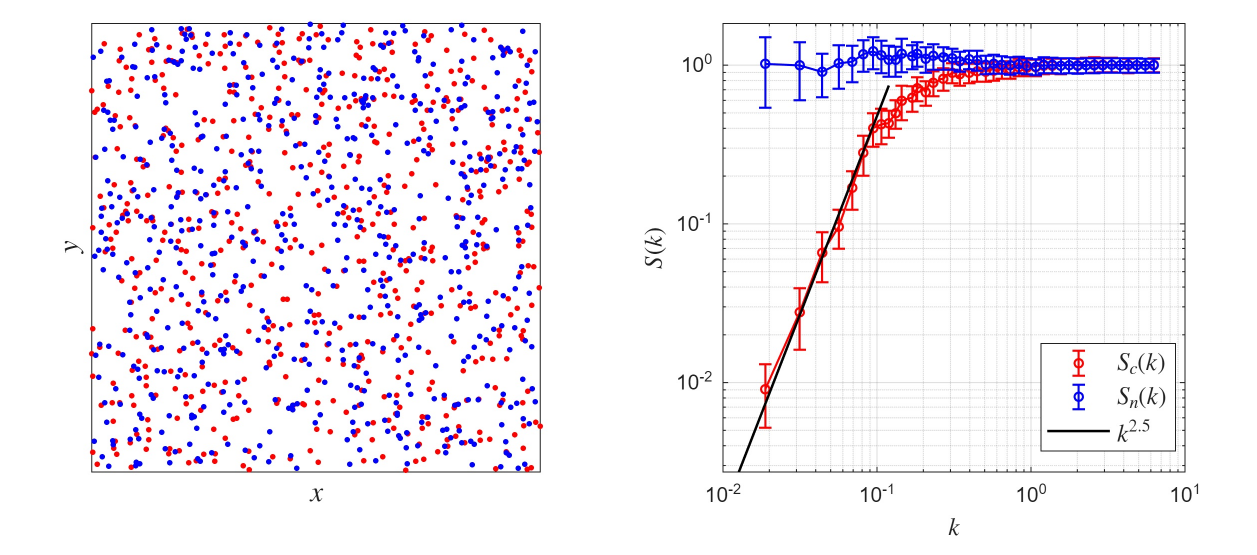}
\includegraphics[width=0.8\textwidth]{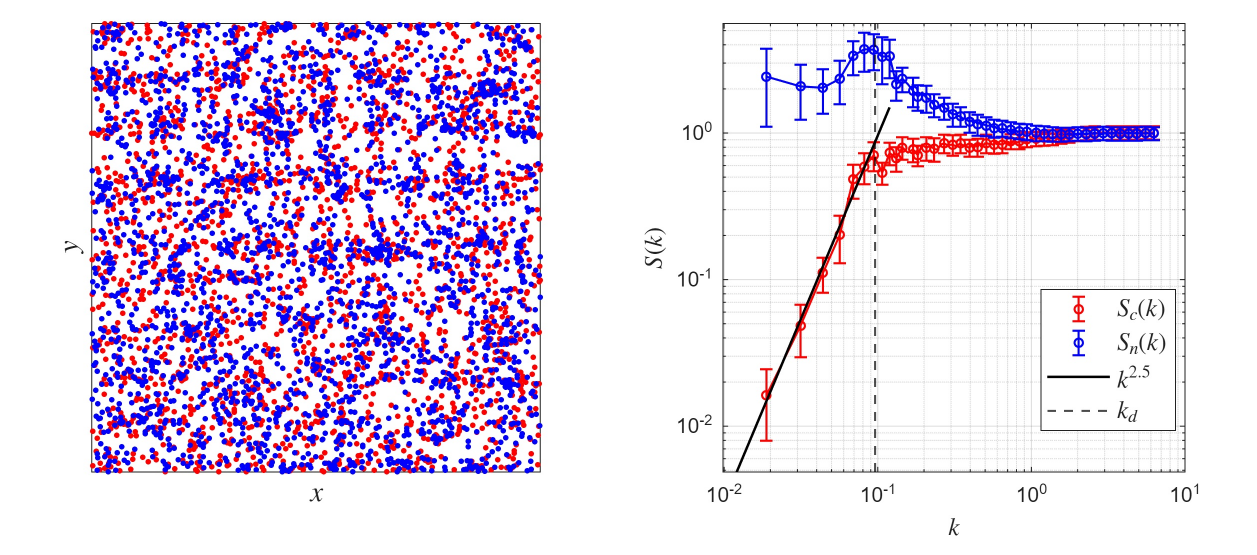}
\caption{Disordered hyperuniformity of the vortex charge distribution with $T/T_c=0.813$, $v^s_y=1.19$ (\textbf{top}) and $T/T_c=0.508$, $v^s_y=4.24$ (\textbf{bottom}). \textbf{Left:} Snapshot of the vortex configuration at $t=2000$; red and blue markers denote positive and negative circulation, respectively. \textbf{Right:} Time-averaged structure factors. The number structure factor $S_n(k)$ (blue) does not vanish as $k\to0$, indicating no long-range number order. The charge structure factor $S_c(k)$ (red) exhibits a clear power-law suppression $S_c(k)\propto k^{\alpha}$ with $\alpha \approx 2.5$ as $k\to0$, demonstrating Class I disordered hyperuniformity. The exponent $\alpha \approx 2.5$ is shown to be robust across different temperatures. Vertical line in bottom right panel corresponds to mean cluster distance $d=2\pi/k_d$ identified by DPC algorithm explained in last section.}
\label{fig:structure_factor2}
\end{figure}

\section{Temperature robustness of hyperuniformity}\label{app:t_dependence}

To verify the universality and temperature robustness of the topological charge hyperuniformity, we analyze the steady-state vortex distributions across different temperatures, specifically for $T/T_c = 0.813$ ($v^s_y = 1.19$) and $T/T_c = 0.508$ ($v^s_y = 4.24$). As shown in Fig.~\ref{fig:structure_factor2}, while the vortex number structure factor $S_n(k)$ remains finite as $k \to 0$, confirming the absence of long-range positional order in vortex numbers, the topological charge structure factor $S_c(k)$ consistently exhibits a robust power-law suppression $S_c(k) \propto k^\alpha$ with a stable exponent $\alpha \approx 2.5 > 1$ across all examined temperatures. This demonstrates that Class I disordered hyperuniformity of topological charges is an intrinsic and universal non-equilibrium feature of strongly coupled holographic superfluid turbulence, independent of temperature and superflow driving strength.

A notable feature in Fig.~\ref{fig:structure_factor2} is the temperature-dependent behavior of the low-wavenumber peak in \(S_n(k)\). For \(T/T_c = 0.813\) (upper panel), no pronounced peak is observed, indicating the absence of macroscopic spatial clustering of vortices at this temperature. In contrast, for \(T/T_c = 0.508\) (lower panel), a clear peak appears at \(k_d \simeq 0.09\), corresponding to a characteristic real-space cluster spacing \(d = 2\pi/k_d \simeq 70\) identified by the DPC algorithm (see Sec.\,S3). This temperature dependence reflects the interplay between the superflow driving strength and the intrinsic dissipation of the holographic superfluid. At higher temperatures (closer to \(T_c\)), the superfluid density is lower, and the stronger dissipation suppresses the formation of long-lived high-density vortex clusters. At lower temperatures, the larger superfluid density and stronger driving sustain more pronounced density modulations. Crucially, despite this temperature-dependent behavior of the number structure factor, the charge structure factor \(S_c(k)\) remains robustly hyperuniform with \(\alpha \approx 2.5\) across all temperatures, confirming that the topological charge ordering is a universal feature of the non-equilibrium steady state, independent of the detailed spatial organization of vortex number density.

\section{Freely decaying vortex turbulence}\label{app:decay}

The study of quantum turbulence within the holographic framework was largely ignited by the seminal work of \cite{chesler2013holographic} and subsequently expanded by \cite{lan2016towards} to alternative quantization. By investigating the freely decaying turbulence of two-dimensional holographic superfluids, these studies reported the emergence of a Kolmogorov-like $k^{-5/3}$ scaling law in the kinetic energy spectrum. However, this conclusion has been heavily scrutinized when compared to the well-established phenomenology of the GPE.

As systematically demonstrated by \cite{bradley2012energy} and further corroborated by \cite{billam2015spectral}, the manifestation of a genuine $k^{-5/3}$ inertial range in two-dimensional quantum turbulence strictly requires an inverse energy cascade. Physically, this inverse cascade is driven by the spatial aggregation of same-sign topological defects into macroscopic vortex clusters (often referred to as Onsager vortices). In stark contrast, holographic superfluid turbulence exhibits a direct energy cascade, where energy flows from large scales down to small scales before being dissipated by the black hole horizon, and conspicuously lacks the formation of such large-scale vortex clusters. Consequently, the underlying physical mechanism required to support a true $-5/3$ scaling is entirely absent in these holographic systems. 

\begin{figure}[htbp]
\centering
\includegraphics[width=1\textwidth]{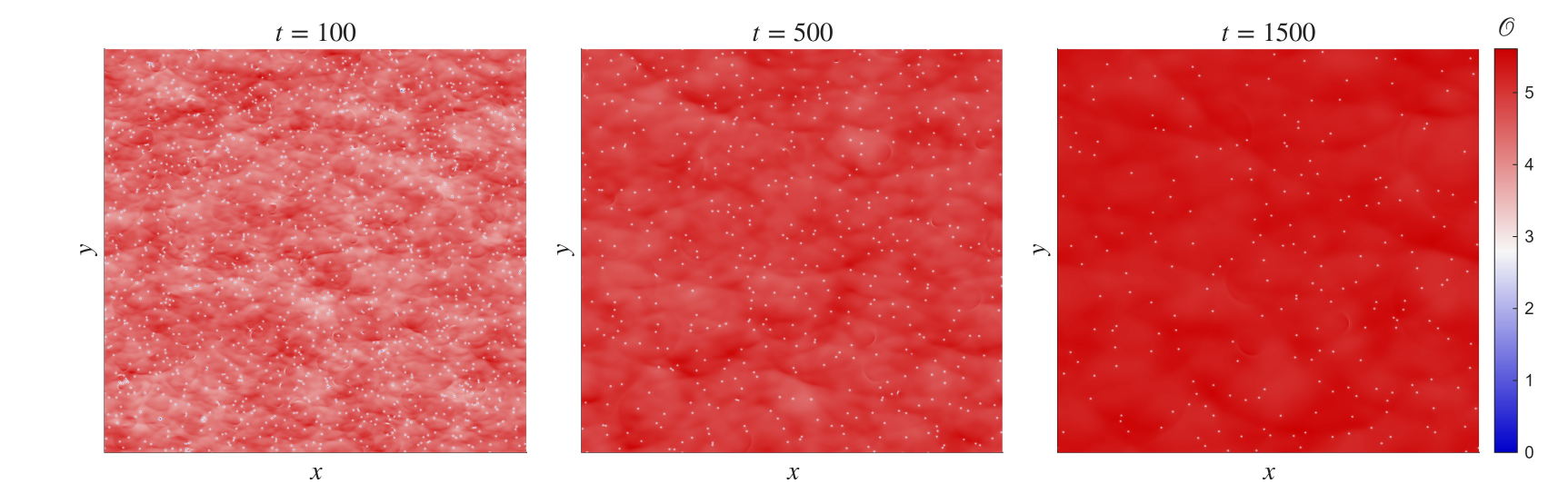}
\includegraphics[width=1\textwidth]{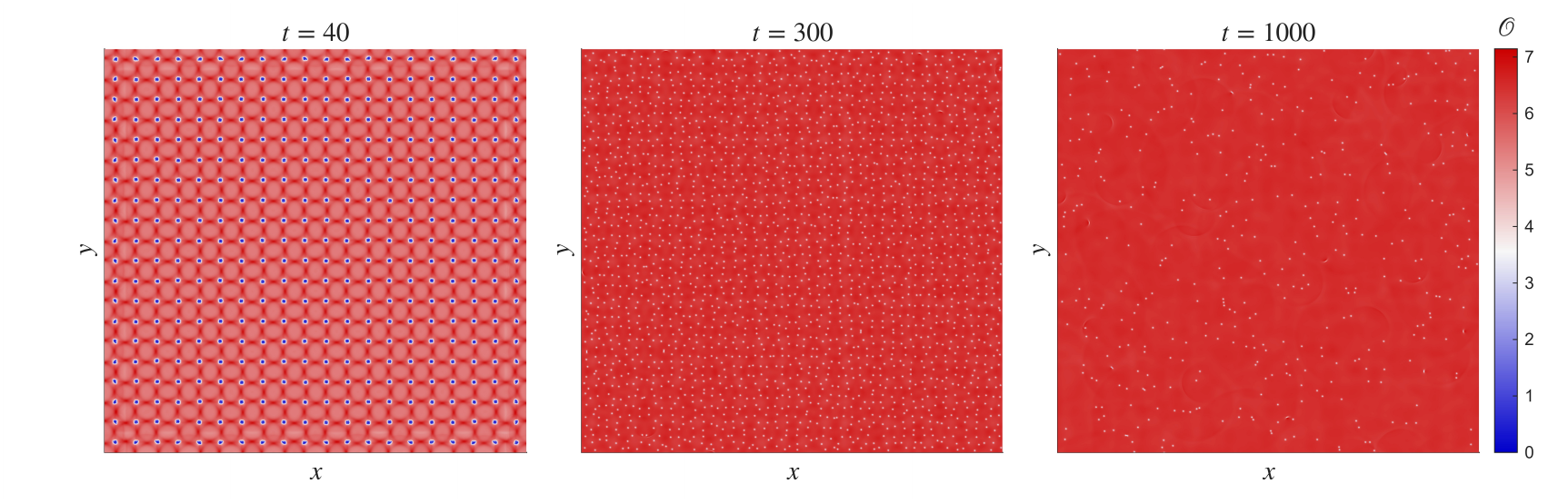}
\caption{Snapshots of order parameter magnitude $\mathcal{O}$ of two types of initial conditions for the freely decaying turbulence. \textbf{Top:} Initially 2500 vortices are randomly distributed. From left to right are snapshots at $t=100$, $t=500$ and $t=1500$. \textbf{Bottom:} Initially vortex lattice with 200 pairs of vortices with winding number $W=\pm6$ are put by hand. From left to right are snapshots at $t=40$, $t=300$ and $t=1000$. Plotted spatial region is $[0,\,500]\times[0,\,500]$.
\label{fig:movie_decay}}
\end{figure}

\begin{figure}[htbp]
\centering
\includegraphics[width=0.45\textwidth]{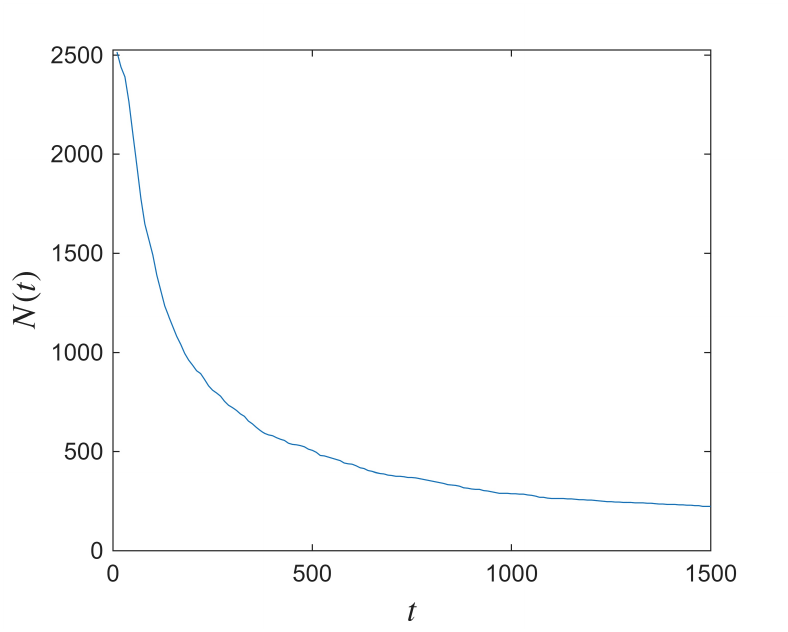}
\includegraphics[width=0.455\textwidth]{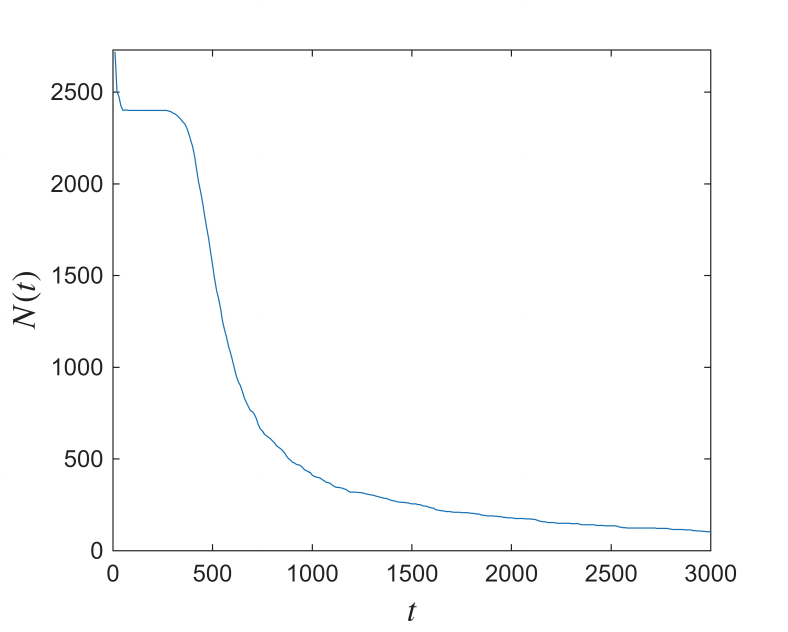}
\caption{Time evolution of the total vortex number $N(t)$ for the freely decaying turbulence. \textbf{Left:} Initial state with randomly distributed vortices. \textbf{Right:} Initial state with a vortex lattice. In both cases, the total vortex number decays rapidly with time, characteristic of decaying turbulence. We choose the same setup as Fig.~\ref{fig:movie_decay}.
\label{fig:NofV_decay}}
\end{figure}

Addressing this discrepancy, the authors of~\cite{billam2015spectral} compellingly argued that the apparent $k^{-5/3}$ spectrum observed in early holographic models is not a genuine inertial range. Rather, it is a smooth crossover between a $k^{-1}$ scaling at intermediate wavenumbers arising from the velocity field of isolated vortex and a $k^{-3}$ scaling at higher wavenumbers associated with vortex core structure. 

In this section, we initialize the system with an ensemble of vortices to study freely decaying turbulence in complementary to driven steady-state setup. To simulate freely decaying turbulence, we begin with a homogeneous and static superfluid background solution. We then manually imprint quantized vortices onto this background by introducing localized phase singularities. To ensure our conclusions are robust against initial configurations, we investigate two distinct initial states: (i) a random distribution of 2500 vortices, and (ii) an ordered vortex lattice consisting of 200 pairs of vortices with winding numbers $W=\pm 6$. To be specific, the initial condition of the scalar field $\Psi$ is 
\begin{equation}
    \Psi=\Psi_b[1-\sum_{k=1}^{N_v}e^{-\frac{(x-x_k)^2+(y-y_k)^2}{\sigma^2}}]e^{iW\sum_{k=1}^{\frac{N_v}{2}}\mathrm{arctan}(\frac{y-y_k}{x-x_k})-iW\sum_{k=\frac{N_v}{2}+1}^{N_v}\mathrm{arctan}(\frac{y-y_k}{x-x_k})},
\end{equation}
where $\Psi_b$ is the background superfluid solution with no vortex, achieved by solving the homogeneous stationary equations of motion. For both cases, the background stationary state is chosen to be $\mu=6$ and $v^s_i=0$. $N_v$ is the total number of vortices, $(x_k,\,y_k)$ are the initial positions of vortices. $W$ is the winding number of imprinted vortices, which is 1 for randomly distributed vortices and 6 for initial vortex lattice. $\sigma$ is chosen to be 0.5 for randomly distributed vortices and 3 for initial vortex lattice. The system is then allowed to evolve freely according to the equations of motion.

The real-time dynamics of these decaying configurations are visualized in Fig.~\ref{fig:movie_decay}, which displays snapshots of the order parameter magnitude $\mathcal{O}$. As the systems evolve, vortex-antivortex pairs undergo frequent annihilations. This is quantitatively supported by Fig.~\ref{fig:NofV_decay}, which demonstrates a rapid and drastic decay in the total vortex number over time for both initial configurations, a hallmark of decaying turbulence.

Helmholtz decomposition of the superfluid velocity field is shown in Fig.~\ref{fig:velocity}. The corresponding energy spectra for the random and lattice initial conditions are presented in Figs.~\ref{fig:spectrum_decay} and~\ref{fig:spectrum_lattice}, respectively. For the case of initially randomly distributed vortices, the total and incompressible energy spectra, $E(k)$ and $E_i(k)$, exhibit a stable $k^{-1}$ scaling at small wavenumbers reflecting the velocity field of isolated vortices and a $k^{-3}$ scaling at large wavenumbers associated with the vortex core structure throughout the entire temporal evolution. Crucially, a $k^{-5/3}$ scaling appears only as a narrow and transient crossover region connecting these two distinct scales, never developing into an extended inertial range. Furthermore, the compressible spectrum $E_c(k)$ initially displays a $k^{-2}$ scaling driven by dense, shock-like structures resulting from intense vortex-antivortex annihilations. As the system evolves and annihilation events become rare, this $k^{-2}$ scaling rapidly vanishes.

\begin{figure}[htbp]
\centering
\includegraphics[width=1\textwidth]{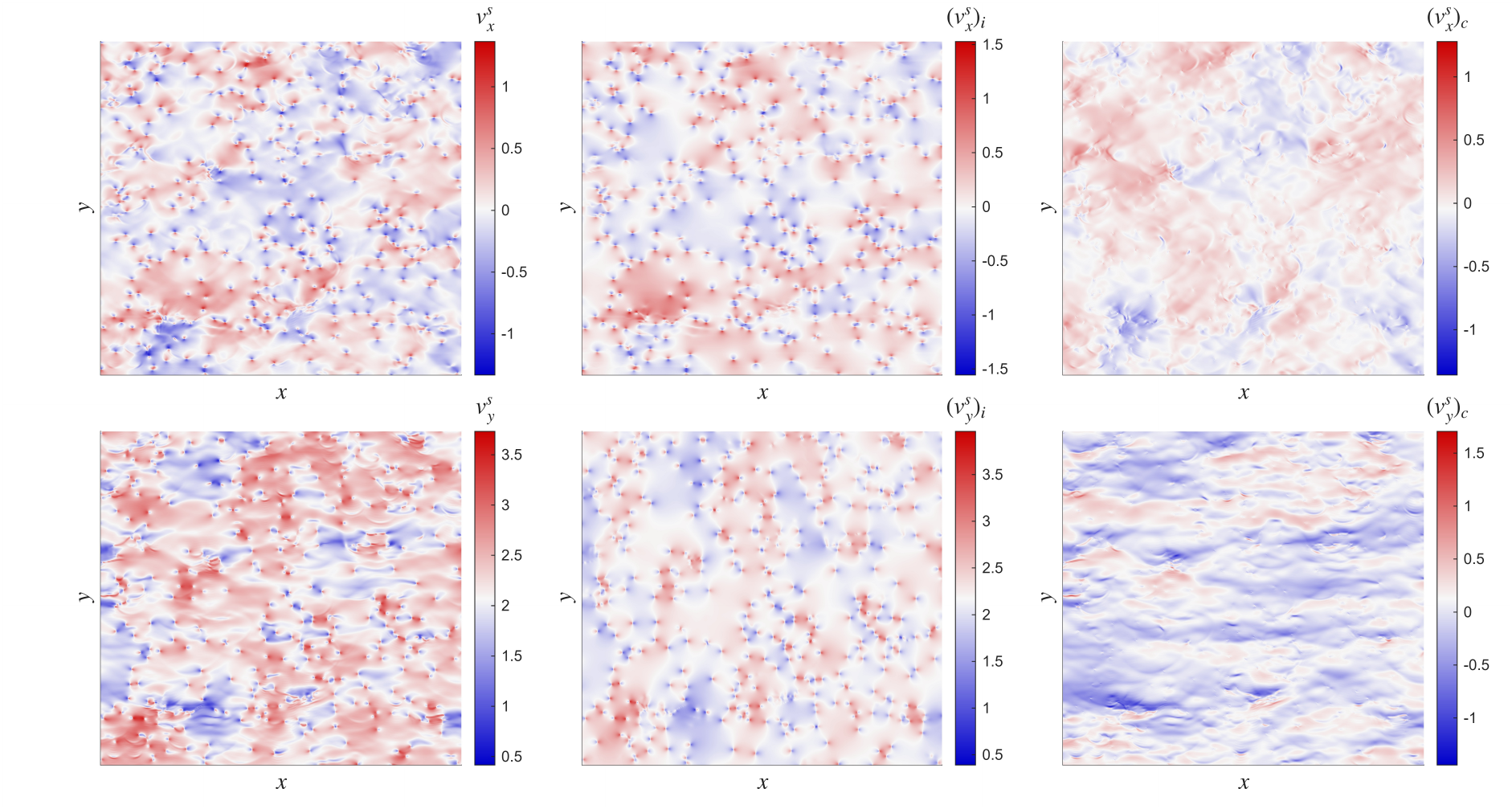}
\caption{Helmholtz decomposition of the superfluid velocity field for the freely decaying turbulence. \textbf{Top:} Total superfluid velocity $v^s_i$. \textbf{Middle:} Incompressible component $(v^s_i)_i$. \textbf{Bottom:} Compressible component $(v^s_i)_c$. The spatial region shown is $[0,\,200]\times[0,\,200]$. The incompressible velocity field is dominated by vortex-induced velocity field, whereas the compressible component is composed of various density perturbations, including sound waves generated by vortex-antivortex annihilation. 
\label{fig:velocity}}
\end{figure}

In contrast, the turbulent evolution originating from a vortex lattice displays a distinct early-time behavior. Because the initial lattice configuration is constructed from highly charged vortex pairs with $W=\pm 6$, these multi-quantized vortices effectively act as macroscopic spatial clusters at the onset of the simulation (see bottom panel of Fig.~\ref{fig:movie_decay}). To quantitatively verify this clustering, we compute the vortex cluster correlation function, defined as $C = \frac{1}{N} \sum_{i=1}^N c_i$, where $N$ is the total vortex number, and $c_i = \pm1$ if the circulation of the nearest neighbor of the $i$-th vortex has the same or opposite sign respectively. At $t=300$, this correlation function evaluates to $C = 0.3076$, providing direct quantitative evidence for the presence of same-sign vortex clusters. Consequently, at these early times such as $t=300$, $E(k)$ and $E_i(k)$ indeed exhibit a discernible and wider $k^{-5/3}$ scaling region (Fig.~\ref{fig:spectrum_lattice}). This observation is highly consistent with conclusions from previous studies \cite{bradley2012energy,billam2015spectral,yang2024emergence}, which established that the spatial aggregation of same-sign vortices is the physical prerequisite for generating a $k^{-5/3}$ spectrum. However, as the highly charged vortices inevitably decay into singly quantized vortices and these clusters dissolve over time, this $k^{-5/3}$ region progressively narrows and is eventually replaced by the $k^{-1}$ scaling. 

Despite their distinct initial configurations and early-time dynamics, the late-time behaviors of both systems converge. In the late stages of decaying turbulence, neither the initially random distribution nor the dissolving vortex lattice sustains a genuine $k^{-5/3}$ inertial range. Ultimately, both scenarios robustly support the conclusion that, in the absence of stable macroscopic vortex clustering, the apparent $k^{-5/3}$ behavior in these holographic superfluids is merely a transient short crossover rather than a true persistent scaling law. As the memory of the initial state fades, the system transitions into a universal late-time regime characterized by a $k^{-1}$ energy spectrum at intermediate scale. A similar dependence of energy spectrum on initial conditions at intermediate times, eventually giving way to a universal late-time spectrum consistent with $k^{-1}$, has also been briefly noted in~\cite{ewerz2015non}. However, the authors of~\cite{ewerz2015non} report a transient $k^{-5/3}$ scaling for random distributions rather than for vortex lattices. This discrepancy may arise from their use of a different definition for the energy spectrum compared to ours.

\begin{figure}[htbp]
\centering
\includegraphics[width=1\textwidth]{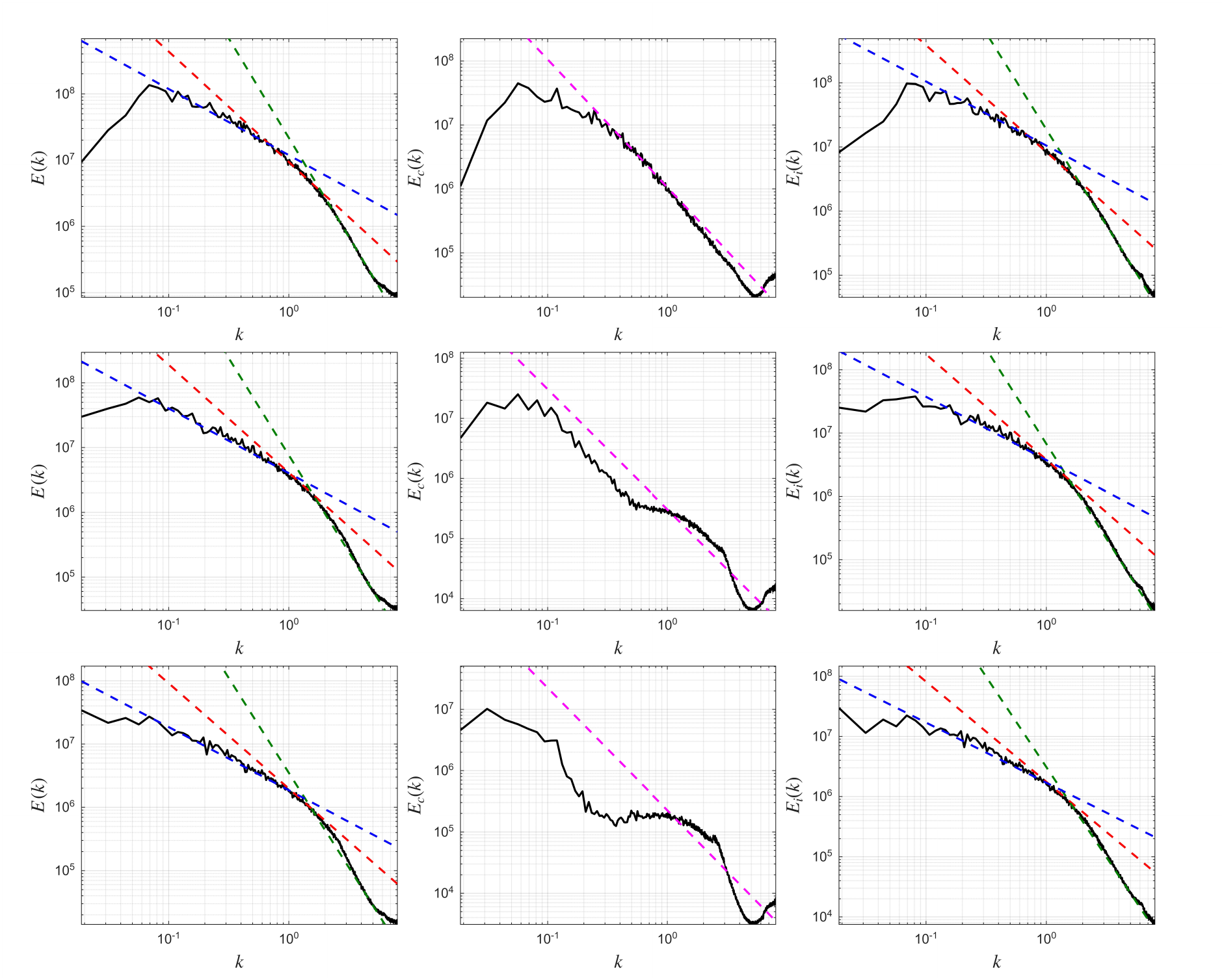}
\caption{Energy spectra of decaying vortex turbulence originating from a random distribution of vortices. The total $E(k)$, compressible $E_c(k)$, and incompressible $E_i(k)$ energy spectra are shown from left to right, with times $t=100$, $t=500$, and $t=1500$ arranged from top to bottom. The dashed lines denote various power laws: Kolmogorov's $k^{-5/3}$ (red), $k^{-1}$ (blue), $k^{-3}$ (green), and $k^{-2}$ (magenta). For all times, $E(k)$ and $E_i(k)$ exhibit a clear $k^{-1}$ scaling at small $k$ and a $k^{-3}$ scaling at large $k$, whereas the $k^{-5/3}$ scaling appears only as a short crossover rather than an extended inertial range. At $t=100$, $E_c(k)$ displays a $k^{-2}$ scaling, which may be attributed to dense shock-like structures induced by vortex-antivortex annihilations. At later times, as annihilation events become much rarer, this $k^{-2}$ scaling disappears.
\label{fig:spectrum_decay}}
\end{figure}

\begin{figure}[htbp]
\centering
\includegraphics[width=1\textwidth]{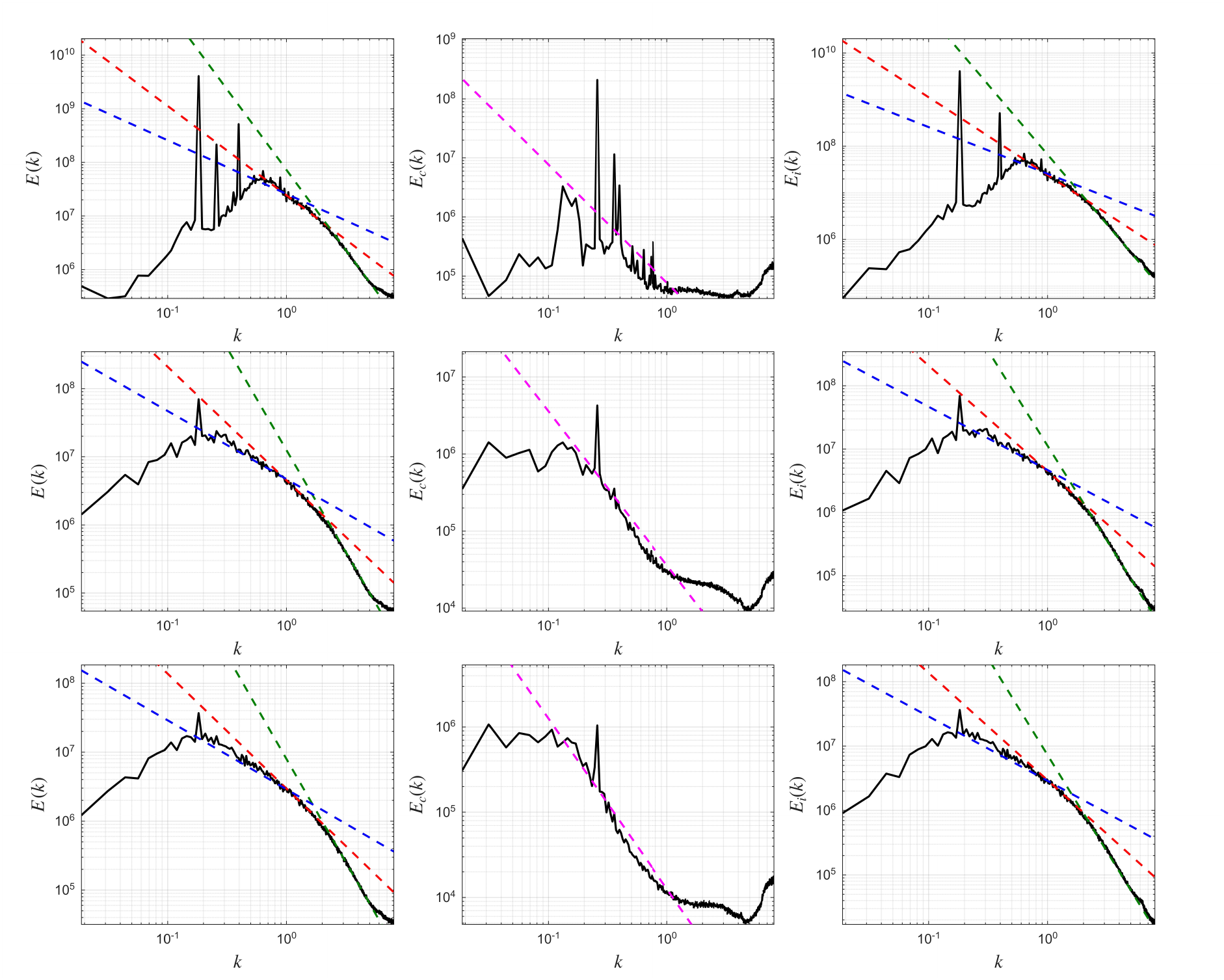}
\caption{Energy spectra of decaying vortex turbulence originating from an initial vortex lattice. The total $E(k)$, compressible $E_c(k)$, and incompressible $E_i(k)$ energy spectra are shown from left to right, with times $t=300$, $t=1000$, and $t=1400$ arranged from top to bottom. The dashed lines indicate reference power laws: Kolmogorov's $k^{-5/3}$ (red), $k^{-1}$ (blue), $k^{-3}$ (green), and $k^{-2}$ (magenta). At $t=300$, $E(k)$ and $E_i(k)$ display a $k^{-5/3}$ scaling region at intermediate $k$; however, this region narrows and is gradually replaced by a $k^{-1}$ scaling as time progresses. At large $k$, a $k^{-3}$ scaling persists throughout the evolution. For $E_c(k)$, no clear scaling region is identified.
\label{fig:spectrum_lattice}}
\end{figure}

\end{document}